%% file: nice_paper.tex
\documentclass[10pt, 5p, nopreprintline]{elsarticle}

\journal{TBA}

\biboptions{numbers,sort&compress}

\usepackage{libertine}
\usepackage{float}
\usepackage[dvipsnames]{xcolor}
\usepackage{libertinust1math}
\usepackage{soul}
\usepackage{amsmath}                 
\usepackage{bbold}                   
\usepackage{graphicx}                
\usepackage{eurosym}                 
\usepackage{mathtools}               
\usepackage{url}                     
\usepackage{booktabs}                
\usepackage{epstopdf}                
\usepackage{xfrac}                   
\usepackage{tabularx}                
\usepackage{bm}                      
\usepackage{subfig}
\usepackage{subcaption}              
\usepackage{longtable}               
\usepackage{multirow}                
\usepackage{threeparttable}          
\usepackage{pdflscape}               
\usepackage[export]{adjustbox}       
\usepackage[version=4]{mhchem}       
\usepackage{xurl} 				     
\usepackage[colorlinks]{hyperref}    
\usepackage[parfill]{parskip}        
\usepackage[nameinlink,sort&compress,capitalise,noabbrev]{cleveref} 
\usepackage[leftcaption,raggedright]{sidecap} 
\usepackage[prependcaption,textsize=footnotesize]{todonotes} 
\usepackage{siunitx}                 
\usepackage{csquotes} 				 
\usepackage[acronym, automake, nonumberlist]{glossaries-extra}
\usepackage{float}
\usepackage{lipsum}
\usepackage{mdframed}
\usepackage{microtype}
\usepackage[resetlabels,labeled]{multibib} 
\newcites{S}{Supplementary References}
\bibliographystyleS{elsarticle-num}

\usepackage{etoolbox}
\makeatletter
\renewenvironment{abstract}{\global\setbox\absbox=\vbox\bgroup
  \hsize=\textwidth%
  {\centering\textbf{\@elsarticleabstitle}\par}%
  \medskip\noindent\unskip\ignorespaces}
  {\egroup}
\patchcmd{\MaketitleBox}{\hrule\vskip12pt}{\vskip12pt}{}{}
\patchcmd{\MaketitleBox}{\hrule\vskip12pt}{\vskip12pt}{}{}
\patchcmd{\MaketitleBox}{\vskip36pt}{\vskip20pt}{}{}
\makeatother

\DeclareSIUnit\year{a}
\DeclareSIUnit{\tco}{t_{\ce{CO2}}}
\DeclareSIUnit{\sieuro}{\mbox{\euro}}

\graphicspath{ 						
	{imgs/},
	}

\def\th{${}_{\textrm{th}}$}
\def\deg{${}^\circ$}

\newdefinition{rmk}{Remark}

\makeglossaries

\newacronym{iea}{IEA}{International Energy Agency}

\begin{document}
\normalsize

\begin{frontmatter}

  \title{Accelerating fossil gas independence in Europe}

	\author[tub_address]{Lukas Franken} \ead{lukas.franken@tu-berlin.de}
	\author[tub_address]{Iegor Riepin}
	\author[tub_address]{Tom Brown}

	\address[tub_address]{Department of Digital Transformation in Energy Systems, Technische Universit\"at Berlin, Germany}

    \begin{abstract}
	   \begin{center}\begin{minipage}{0.82\textwidth}
	   \input{sections/abstract.tex}
	   \end{minipage}\end{center}


        \par\vspace{4.0em}
        \centerline{\includegraphics[width=\textwidth, height=0.52\textheight, keepaspectratio]{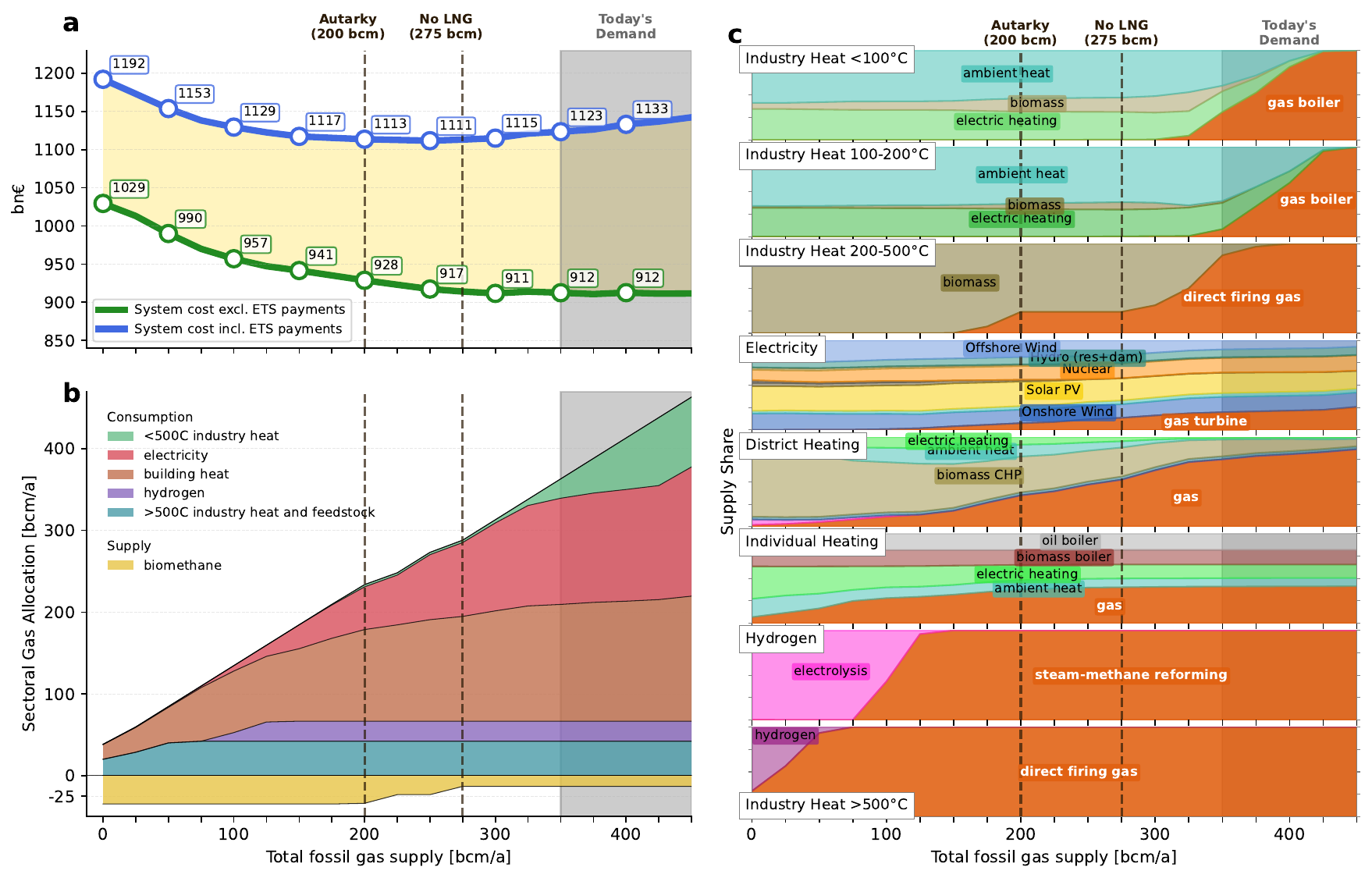}}
	\end{abstract}

\end{frontmatter}

\clearpage

\section*{Introduction}
\input{sections/introduction}

\section*{Cost-Optimal Gas Phase-Out is Concentrated in Low-Temperature Industry Heat and Bulk Power}

\input{sections/results1.tex}

\section*{Heat Electrification is the Binding Constraint on LNG-Independence Before 2040}
\input{sections/results4.tex}

\section*{At a Carbon Price of 100\,\euro/tCO2, Autarky is Cost-Optimal from a Gas Price of 30\,\euro/MWh}
\input{sections/results3.tex}

\section*{Gas Continues to Anchor Electricity Marginal Prices Long After it Leaves the Bulk Power Mix}
\input{sections/results2.tex}
\section*{Consumer Exposure to the Global Gas Price under Present Gas Use Levels}
\input{sections/results5.tex}

\section*{Discussion}
\input{sections/discussion.tex}

\section*{Methods}
\input{sections/methods.tex}

\section*{CRediT authorship contribution statement}
\textbf{Lukas Franken:} Writing – original draft, Writing – review \& editing, Visualization, Validation, Software, Methodology, Investigation, Formal analysis, Conceptualization.
\textbf{Iegor Riepin:} Writing – review \& editing, Conceptualization, Methodology, Project administration.
\textbf{Tom Brown:} Writing – review \& editing, Conceptualization, Methodology, Funding acquisition.

\section*{Declaration of Competing Interests}
The authors declare no competing interests.

\section*{Acknowledgements}
We thank Bobby Xiong, Anne Neumann, Markus Millinger, Fabian Neumann, Tobias Fleiter, Khaled Al-Dabbas and Iris Rieth-Menze for fruitful discussions.

\section*{Funding Statements}
L.F. and I.R. acknowledge support by the Federal Ministry for Economic Affairs and Energy of Germany (BMWE) jointly with the \href{https://cetpartnership.eu/}{CETPartnership} through the Joint Call 2022. As such, L.F. and I.R. further acknowledge funding from the European Union's Horizon Europe research and innovation programme under grant agreement no. 101069750.

\section*{Data availability}
The open source code and data to reproduce our experiments, including the download and installation of data and software dependencies, are available on GitHub under the MIT license at \url{https://github.com/LukasFrankenQ/resilient-pypsa-eur}.


\addcontentsline{toc}{section}{References}
\renewcommand{\ttdefault}{\sfdefault}
\bibliography{refs}

\section*{Appendix}
\input{sections/appendix.tex}

\makeatletter
\renewcommand \thesection{S\@arabic\c@section}
\renewcommand\thetable{S\@arabic\c@table}
\renewcommand \thefigure{S\@arabic\c@figure}
\makeatother
\renewcommand{\citenumfont}[1]{S#1}
\setcounter{equation}{0}
\setcounter{figure}{0}
\setcounter{table}{0}
\setcounter{section}{0}

\end{document}

%% file: sections/abstract.tex
Recent price shocks have prompted calls to curb Europe's dependence on fossil gas imports, but the cost of this goal, and the consumer protection it affords, remain uncertain.
Here we address this gap by imposing constraints on fossil gas supply in a European energy system model that co-optimises abatement across all gas uses at high spatio-temporal resolution.
Cutting import reliance proves economically compelling:
through savings in power generation and low-temperature heat in industry and buildings, Europe can halve its natural gas consumption for 16bn\euro/a, aligning demand with the continent's production capacity of 200$\,$bcm.
This extra system cost is comparable to what consumers spend today on a 2$\,$\euro/MWh rise in gas import prices.
However, this sovereignty alone does not shield consumers from global gas price volatility:
we find that, even at a small share of the mix, gas remains dominant in shaping the marginal electricity price, leaving consumers exposed without additional policy measures.


%% file: sections/introduction.tex
Gas is deeply embedded in European economies.
Over the last ten years, annual consumption has hovered between 350 and 450 bcm, making up around a quarter of Europe's total primary energy demand \cite{eurostat_energy_balances_2024, beis_ecuk_2021}.
Half of this consumption, around 200$\,$bcm, can be produced domestically; for the remainder, Europe relies on imports \cite{norskpetroleum_exports_2024}.

\begin{figure*}
    \centering
    \includegraphics[width=1\linewidth]{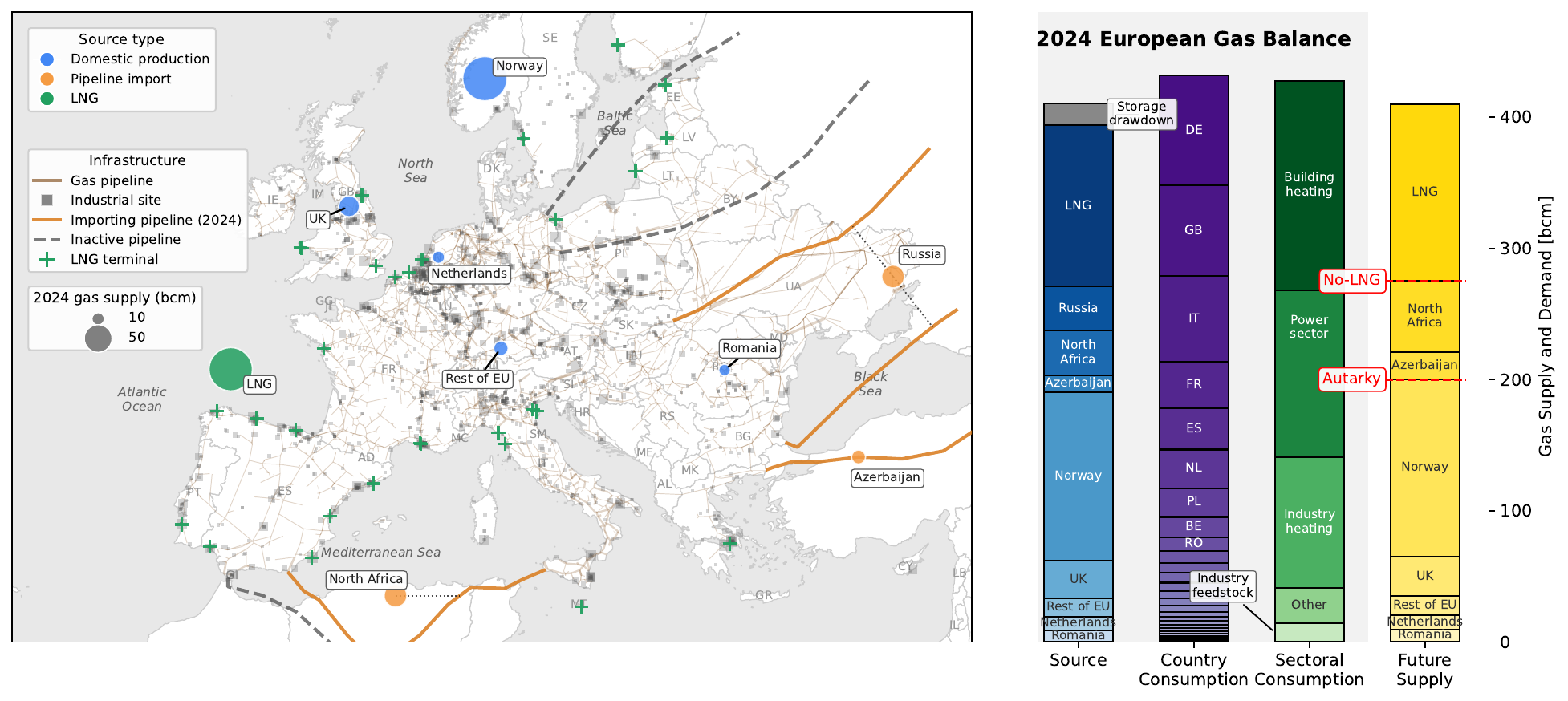}
    \caption{
      \textbf{a} Map of European gas infrastructure and supply pathways. \textbf{b} Supply and consumption split by country and sector. Consumption data from Eurostat \cite{eurostat_gasm_2024, eurostat_balc_2024}; supply data from \cite{oies_quarterly_gas_q29_2024}; industrial site locations from \cite{hotmaps_industrial_database, openstreetmap}.
  }
    \label{fig:intro_plot}
\end{figure*}

These gas imports have become an increasingly acute economic risk for Europe.
In 2022, following the near-complete decoupling from Russian fossil fuels, the ensuing scarcity and a pivot to shipped Liquefied Natural Gas (LNG), Europe spent an additional trillion euros on oil, gas and coal relative to 2021 \cite{boccara2023balancing}.
In early 2026, gas imports re-emerged as a vulnerability when the closure of the Strait of Hormuz tightened global LNG supply by 20\% and European gas prices roughly doubled to 45--60\euro/MWh \cite{tradingeconomics_ttf_2026}.
Two shocks within five years have triggered a debate in Europe about ending gas import reliance altogether \cite{graichen2026gasabhaengigkeit, gillot2026, rosenow2026war};
initially considered a safe alternative, the new LNG dependence reinforces the coupling between volatile global gas markets and consumer prices across the broad spectrum of gas applications \cite{zwickl_bernhard_hormuz_2026} (Fig \ref{fig:intro_plot}).

In European policy, the drive to lower gas reliance doubles down on existing decarbonisation efforts.
In 2022, the \textit{RePowerEU} package allocated an additional 290$\,$bn\euro$\,$to medium- and long-term gas demand reduction \cite{EuropeanCommission2022REPowerEU}, on top of the 2021 \textit{Fit-for-55} package, which already targeted a 30\% reduction in gas use by 2030 \cite{EC2021FitFor55}.
As recently as early 2026, the EU decided to end all Russian gas imports by the end of the same year \cite{EC2025RussianGasAgreement, Harneys2026RussianGasOilPhaseout}.
The same communication also includes a directive for member states mandating national plans to reduce fossil gas consumption. 
In response to the closure of the Strait of Hormuz, in April 2026 the EU introduced the \textit{AccelerateEU} programme, which sets more ambitious electrification targets, aims to speed up grid reinforcement, and seeks to protect consumers and industry from energy-price spikes \cite{ec_accelerateeu_2026}.
Meanwhile, the ongoing gas displacement in Europe is mainly driven by decarbonisation measures such as carbon pricing \cite{euets_directive_2023_959, uketsorder2020}, renewable-power subsidies \cite{ceer2025res}, and incentives for households to electrify their heating \cite{ehpa2023subsidies}.

So far, the gas phase-out has progressed unevenly across power, building heat and industry because the cost structure of alternatives varies \cite{iea_gas_2022_drivers, madeddu2020}.
Wind and solar are directly displacing bulk power generation \cite{ember_eer_2025}, and batteries are competing with gas to balance an increasingly intermittent generation fleet \cite{iea_electricity_2026}.
In the building sector, Europe currently installs around 2.5 million heat-pump units annually \cite{ehpa_market_report_2025}, mostly displacing gas boilers, though heat pumps still provide only a minority of overall heat \cite{iea_future_heat_pumps_2022}.
Phase-out in industry is more complex.
Most industrial gas is burned for heat, more than two-thirds for temperatures below 500\deg C, where electrification or biomass options are widely considered commercially available \cite{fraunhoferISI2024electrification}.
Higher-temperature demands, such as cement or glass production, are considered hard-to-abate, with expensive and scarce hydrogen as the main alternative \cite{odenweller2025green}.
Hydrogen production itself, along with methanol synthesis, hydrocracking and parts of petrochemical production, relies on gas as a feedstock \cite{iea_global_hydrogen_review_2024, irena_methanol_2021, eia_refinery_hydrogen_2015}; there too gas is hard to replace \cite{iea_future_petrochemicals_2018}.
A final option is to supplement supply with biomethane, produced from maize, manure and agricultural residues: by 2030 the EU targets around 35 bcm of annual production \cite{EuropeanCommission2022REPowerEU}.
However, research suggests that this target exceeds plausible capacity by roughly a factor of two \cite{Fehrenbach2022}.

The literature emphasises the role of ``low-hanging fruit'' in lowering Europe's import dependence \cite{agora_gas_exit_2023, AgoraIndustry2026ElectrifyingHeat}.
These are renewable power paired with the electrification of building heat and low-temperature industry heat, a goal largely aligned with EU policy \cite{EuropeanCommission2022REPowerEU}.
This alignment between decarbonisation and energy-independence targets has been noted \cite{pedersen2022long}, but other work highlights that gas phase-out follows from a decarbonisation drive only under a carbon-pricing regime \cite{jewell2016comparison}; otherwise, import reduction pulls domestic fossils such as lignite back into the market, as for instance in Germany in 2022 \cite{FraunhoferISE2022}.
For hard-to-abate sectors, EU policy allocates substantial funds to electrolysis to scale up production before 2030 \cite{EuropeanCommission2022REPowerEU}, accelerating the deployment of gas substitutes for high-temperature industry.
However, an alternative pathway proposed by \textit{Agora} defers hydrogen uptake to 2035 \cite{agora_gas_exit_2023, AgoraIndustry2026ElectrifyingHeat}.

Existing studies tend to embed analyses of gas phase-out within pathways to net-zero emissions by 2050 \cite{EuropeanCommission2022REPowerEU, pedersen2022long, agora_gas_exit_2023}.
This is because the promise of LNG supply in 2022 left climate goals as the primary motivator for further gas reduction \cite{pedersen2022long}.
By 2026 this perspective has shifted.
The substantial cost volatility from exposure to global LNG markets, together with wider geopolitical uncertainty \cite{graichen2026gasabhaengigkeit}, has elevated supply security and resilience as central objectives.
This raises the question of how much gas demand can be reduced to align either with European production capacity alone (around 200 bcm) \cite{norskpetroleum_exports_2024}, requiring a halving of current consumption, or with combined domestic production and pipeline imports from North Africa and Azerbaijan (around 275 bcm) \cite{hafner2023italy, tapag2026operations}, requiring a reduction of roughly one third.
These are non-trivial yet viable near- to medium-term targets with the potential to meaningfully curb Europe's import vulnerability \cite{gillot2026}.
Existing work fails to isolate the costs, price changes and interventions needed for Europe to reach these milestones.

Much of our understanding of opportunities to reduce natural gas consumption comes from case studies.
These assess the business case for localised alternatives \cite{fraunhoferISI2024electrification, AgoraIndustry2026ElectrifyingHeat, Pastore2022} or model sectors in sequence \cite{agora_gas_exit_2023}.
However, gas supplies around 20\% of European energy \cite{eurostat_energy_balances_2024}, and reducing its consumption is a non-marginal intervention. 
The partial equilibrium perspective is only broken once all coupled sectors are cleared within a single market simulation.

Large-scale sector-coupled energy system models (ESMs) solve this issue, but lack the fidelity of case studies.
This is partly for historical reasons; modern ESMs emerged from power sector applications first \cite{horsch2018pypsa}, gradually expanded to other sectors, such as heat and transport \cite{brown2018synergies}, but the industry sector has remained a holdout, treated exogenously \cite{neumann2023potential}.
As a result, even recent ESM studies fix industry gas phase-out exogenously \cite{pedersen2022long}, and fail to capture how gas scarcity drives up its value to the system, and hence its consumer price.
Capturing this scarcity rent is a key strength of ESMs, but it can only realistically form if the model internalises the cost of gas alternatives in hard-to-abate sectors such as high-temperature industry.
With exogenous industry, the model is left to navigate gas phase-out from power and building sectors, where gas alternatives have a different cost structure, thereby distorting the optimal sectoral ordering of gas consumption reduction.
Meanwhile, integrated assessment models (IAMs) typically have strong sectoral coverage but are built on mathematics that complicate the derivation of scarcity rent \cite{nikas2024three}, and, like some ESM studies \cite{EuropeanCommission2022REPowerEU}, also lack the spatio-temporal resolution crucial to simulating commodity price drivers.

\begin{figure*}[h]
    \centering
    \includegraphics[width=\textwidth]{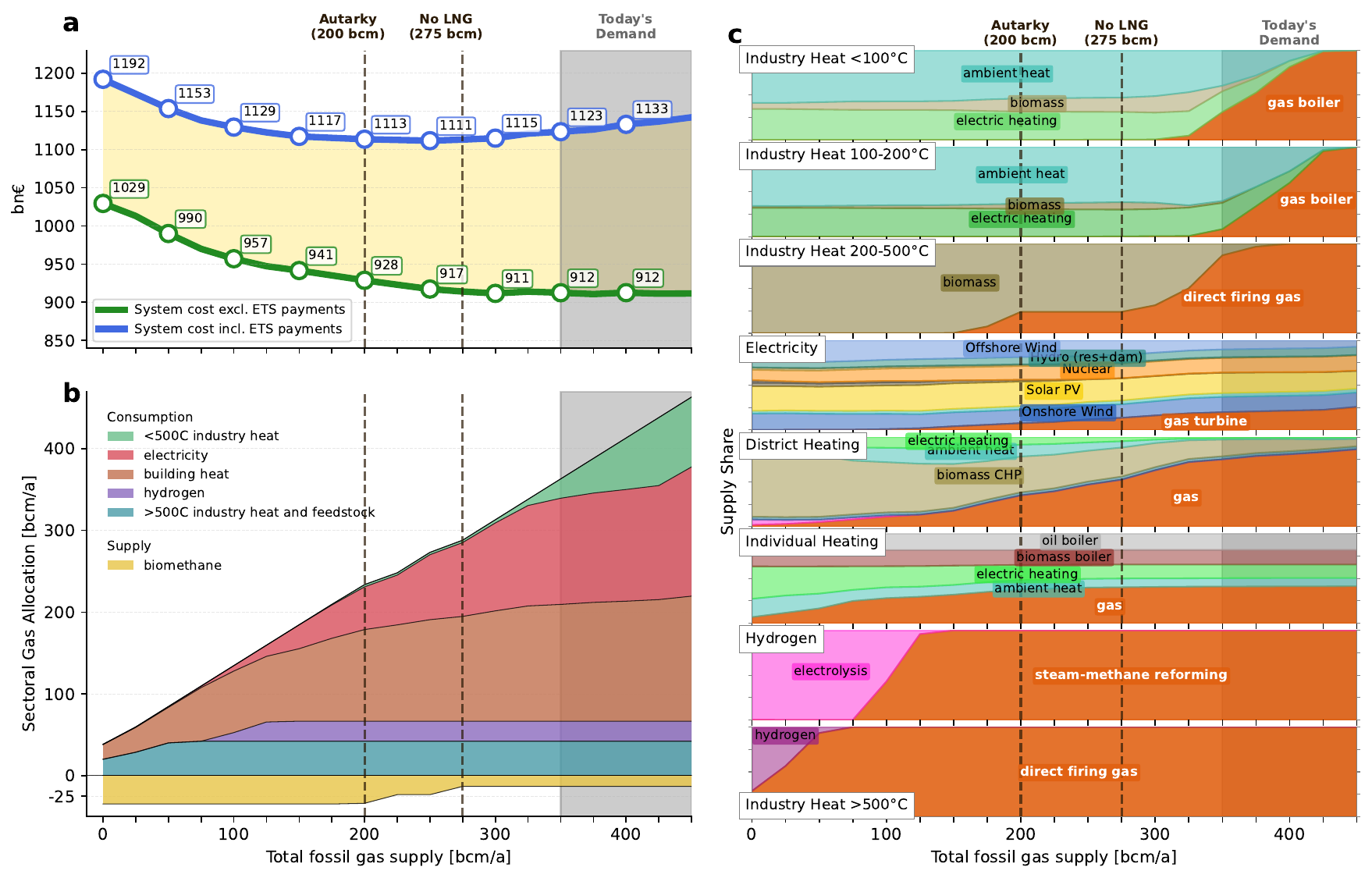}
    \caption{
      European energy system transformation pathways as annual fossil gas supply is constrained to different values ($x$-axis).
      A fixed carbon price of 100 \euro/tCO2 is assumed throughout.
      The vertical shaded areas indicate consumption levels between 2020 and 2025 (Today's Demand), vertical lines highlight the European annual domestic production capacity of 200$\,$bcm (\textit{Autarky} scenario) and domestic production plus pipeline imports from Azerbaijan and North Africa to 275$\,$bcm (\textit{No-LNG} scenario). \textbf{a} Total system cost with and without spending in carbon markets.
      \textbf{b} Gas allocation by sector. Industry heat $<$500°C groups the three heating bands $<$100°C, 100--200°C and 200--500°C, building heat includes residential and service sector heat for rural, individual urban and district demands. Steam-methane reforming with carbon capture is not enabled in the model.
      \textbf{c} Shares of supply mix by sector.
      Note that the slice of industry heat demand shown here refers only to the share currently supplied by gas.
      Further note that the model natively works in TWh. 
      To represent gas-related quantities in billion cubic meters (bcm), we chose a unit conversion factor of 10 as conventions differ around this value.
    }
    \label{fig:results_1_figure}
\end{figure*}

This paper presents the first European energy-system model to optimise gas allocation across all sectors at high spatio-temporal resolution for varying gas consumption levels.
We achieve this by iteratively tightening a cap on fossil gas supply -- from 450$\,$bcm, approximating current European consumption, down to zero in 25$\,$bcm increments.
We use the PyPSA-Eur model, which captures sectoral, geographic and temporal detail in continent-wide energy-economic trade-offs.
The model is further enhanced such that almost all gas uses are fully endogenously represented, including industrial heat.
That means decisions on fuel switching, infrastructure investment, and heat provision route selection are not imposed but emerge as part of the cost-optimisation process, driven by the relative costs across competing supply and substitution options.
As a result, the model can trace a cost-optimal pathway for gas phase-out--revealing the sectoral and regional ordering in which it is most cheaply displaced and the cost penalties incurred when cheap alternatives dry up and the cost of gas abatement rises.

We anchor the analysis to two near- to mid-term gas-consumption milestones that could be achieved in the period 2030--35.
The first is the \textit{No-LNG} scenario with 275$\,$bcm consumption that restricts imports to pipeline gas;
the second is the \textit{Autarky} scenario at 200$\,$bcm that only uses domestic European production.
We compare the cost-optimal roll-out needed to reach the milestones with the current pace of European technology deployment, and quantify the acceleration required for gas-autarkic supply by 2030 or 2035.
We also assess how the commodity gas price--such as the spike caused by the 2026 supply disruptions--influences the gas-consumption level that long-term market equilibrium steers towards.
Finally, the analysis highlights and characterises how gas prices diffuse through the energy system and shape marginal electricity prices, even after gas has largely left the bulk electricity mix.
This enables us to asssess the conditions and policy levers that help to ensure lower gas imports turns into an effective means to protect consumers from spikes in global gas prices.

%% file: sections/results1.tex
We assess total system cost and sectoral implications for different levels of gas consumption.
The market price for gas is set at 24.6$\,$\euro/MWh, representing a uniform commodity price for both domestically produced and imported gas and reflecting the typical post-2022 price regime \cite{tradingeconomics_ttf_2026}.
We frame the analysis as a perspective on 2030--35 and choose the techno-economic and roll-out assumptions, along with a carbon price of 100$\,$\euro/tCO2, accordingly.
If gas consumption stays at today's levels of 350--450$\,$bcm/a, its allocation matches how gas is used in the mid-2020s system--roughly split into thirds between the power sector, building heat and industry (Fig \ref{fig:results_1_figure}).
By inserting the existing fleet of residential heating, we ensure that the split of heat supply across electric, gas, oil and biomass sources is broadly aligned with the real 2020s system (Fig \ref{fig:heat_validation}).

We find that tightening the cap on fossil gas supply leaves system cost largely unchanged deep into the phase-out (Fig \ref{fig:results_1_figure}\textbf{a}).
Indeed, the change in system cost is minimal between the \textit{No-LNG} (275$\,$bcm/a) and \textit{Autarky} (200$\,$bcm/a) scenarios, and rises by up to $\sim$100$\,$bn\euro\ for deeper gas phase-out.
Not accounting for carbon spending, costs increase by $\sim$5$\,$bn\euro/a ($<$1\%) for the \textit{No-LNG} scenario and $\sim$16$\,$bn\euro/a ($<$2\%) for \textit{Autarky}.

Cost-optimal curbing of gas use is largely enabled by phasing out gas in industry heat below 500\deg C and sharply reducing power-sector gas use (Fig \ref{fig:results_1_figure}).
The vast majority of gas is replaced by renewable power generation, heat pumps and biomass boilers.
Biomass use intensifies in the 200--500°C band, where electric alternatives are largely unavailable \cite{AgoraIndustry2026ElectrifyingHeat}.
This pushes biomass consumption from around 1,300 TWh to roughly 1,700 TWh under deeper gas reductions (Fig \ref{fig:total_biomass}; for reference, 1,300$\,$TWh of natural gas corresponds to around 130$\,$bcm).
This level is high and, as in the 2020s, includes unsustainable crop-based sources as well as more sustainable wastes and residues (for reference, 2021 solid biomass use was around 1,300$\,$TWh/a \cite{millinger2025diversity}).
Its allocation, however, partially aligns with cost-optimal biomass use in a future carbon-neutral European energy system, where biomass burning (as it is used here) would be paired with carbon capture to leverage biogenic carbon for negative emissions and renewable fuels \cite{millinger2025diversity}.
The solid-biomass supply curve used in the present model is shown in Fig \ref{fig:biomass_supply_curve}.
There are potential options to electrify industry heat demands above 500°C \cite{glaum2026minimal}, but these are less mature and not included in the present work.

In the power sector, gas gradually exits bulk electricity generation at around 250 bcm of total gas use (Fig \ref{fig:results_1_figure} \textbf{b,c}), but continues to contribute minor volumes well into the phase-out.
This is because gas serves three distinct roles in the power sector, each with different displacement economics.
First, in bulk power generation its competitors are wind and solar, which are directly cost-competitive.
Second, in day-to-day balancing of renewables, batteries displace gas turbines.
Third, backup generation during extended lulls in renewable generation (Dunkelflaute), which batteries would be too expensive to bridge.
Therefore, even at extremely low fossil gas supply, the model does not displace gas-based backup, indicated by generation peaks on a small subset of days (Fig \ref{fig:battery_gas_tradeoff}).

Residential and service-sector heating (hereafter building heat) is the third sector to phase out gas, after industry heat below 500°C and power.
In contrast to the other sectors, where gas phase-out exhibits a sharp ``switching'' behaviour, building heat phases out more gradually (Fig \ref{fig:results_1_figure} \textbf{b} and \textbf{c}).
This is caused by regional differences in the cost of heat when supplied by heat pumps. 
Renewable-power costs and heat-pump coefficients of performance both depend on local conditions, creating regional hold-outs where alternatives to gas are less competitive.
Some modelling choices also contribute:
biomass- and oil-boiler capacity expansion is prohibited, reflecting both their health impact \cite{vicente2018particulate} and ongoing trends \cite{altermatt2023replacing}.
Further, heat supply is forced to be \textit{load-following} (see Methods \ref{subsec:building_heat}), which drives up the levelised cost of heat to 100--200$\,$\euro/MWh, a range broadly in line with the literature \cite{iea_lcoh_chart_2021}.
As a result, even when gas consumption falls below 50 bcm/a, around half is used for residential heating (Fig \ref{fig:results_1_figure}\textbf{a}).
As will be discussed, these results are sound within the logic of the model, but rest on the model's cost assumptions, which can differ from those faced by real consumers.
Real supply choices depend on gas and electricity retail prices, which include grid tariffs, taxes and utility margins that are only partially represented within the model.

Between the \textit{No-LNG} and \textit{Autarky} scenarios (200--275$\,$bcm), the model turns to biogenic methane to compensate for low fossil gas availability.
First, it produces biomethane from straw and sewage sludge (Fig \ref{fig:total_biomass}).
This implies an extensive biomethane build-out of more than 30 bcm, aligned with the RePowerEU target for 2030 (around 35 bcm).
Only after exhausting the biomethane potential does the model replace steam-methane reforming with electrolysis and scale up hydrogen production to displace gas in high-temperature industry applications (Fig \ref{fig:results_1_figure}\textbf{a}).
As gas consumption falls, carbon emissions decline nearly linearly, from roughly 2.3\,GtCO\textsubscript{2} at 450 bcm to 1.8\,GtCO\textsubscript{2} at zero gas use (Fig \ref{fig:co2-balances}).

%% file: sections/results4.tex
\begin{figure*}[t]
    \centering
    \includegraphics[width=\textwidth]{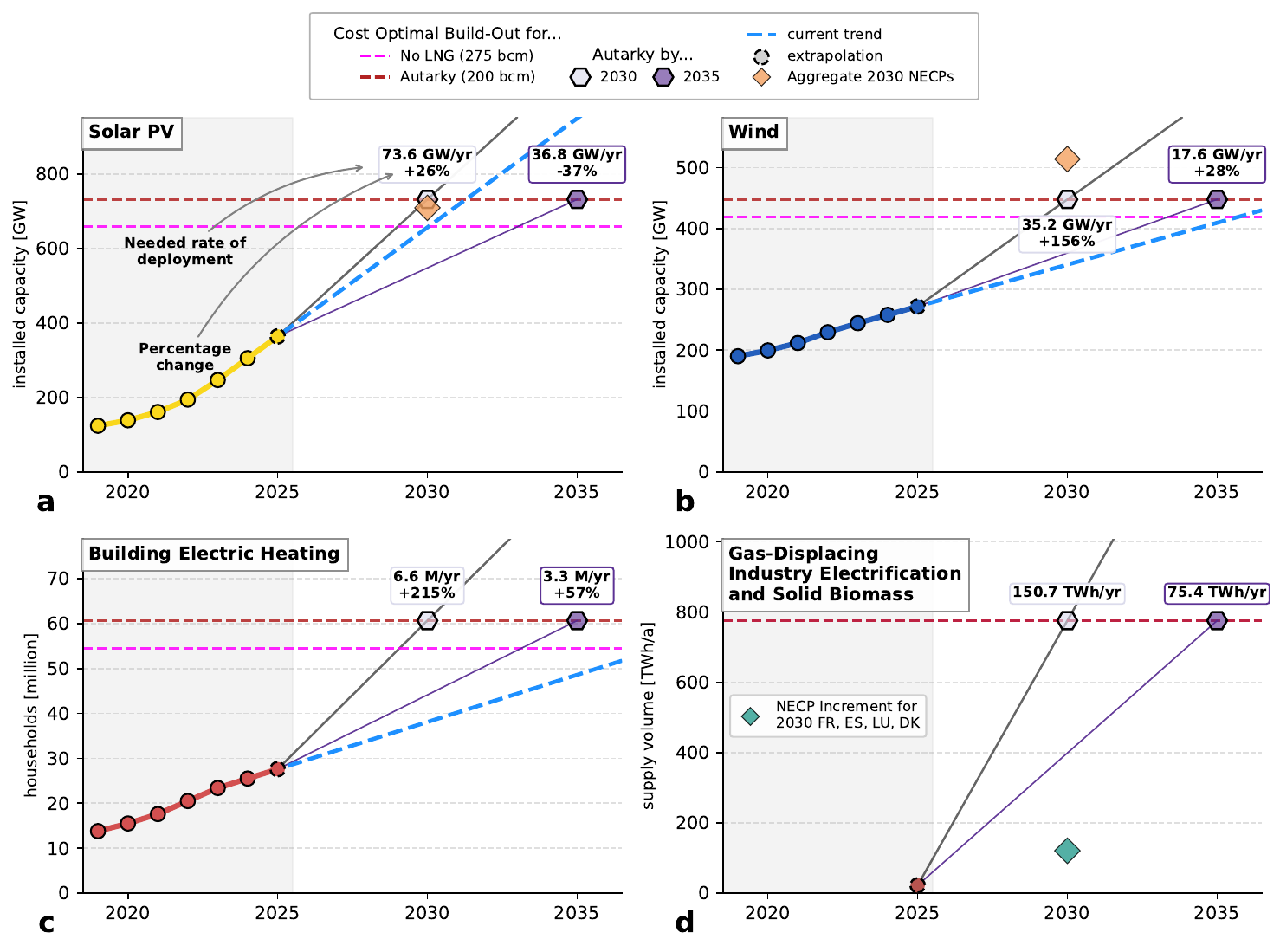}
    \caption{
      Technology roll-out in cost-optimal \textit{No-LNG} and \textit{Autarky} scenarios relative to early 2020s uptake trends.
      Wind combines both on- and offshore installation.
      Note that solid biomass and electric options already supply heat to industry.
      The present figure refers only to the additional heat supply needed to displace demand currently met by gas.
      In both cases, no consistent trend has been observed over the past years.
      National Energy and Climate Plan (NECP) targets for 2030 are from \cite{ember2024necp, desnz2024cleanpower2030}.
      For 2030 projections of gas-displacing industry electricity and biomass usage, we compare 2020 versus 2030 sector usage in \cite{ember2024necp}.
      Both numbers are only reported for FR, ES, LU, DK and so underrepresent real continent-wide expansion plans.
    }
    \label{fig:rollout_trajectory}
\end{figure*}

So far, we have characterised the cost-optimal technology mix at varying fossil gas supply budgets.
However, Europe is already phasing out gas, and incentives to that effect are in place across all relevant sectors --- power, buildings and industry heating.
This raises the question: following current trajectories, when would Europe become LNG-independent (275$\,$bcm) or fully autarkic (200$\,$bcm)?

We find that, relative to 2025 installations, the cost-optimal autarkic scenario requires a doubling of the wind, solar and heat-pump fleets, as well as the displacement of around 80$\,$bcm of gas burned for industry heat below 500°C.
In the power sector, current deployment trends indicate that renewable generation will reach the levels required for gas autarky between 2030 and 2040 (Fig \ref{fig:rollout_trajectory}\textbf{a,b}).
If countries deliver on their National Energy and Climate Plan targets for 2030, this milestone would be reached even sooner.

Gas phase-out in buildings and industry lags behind the power sector.
Current building heat-pump trajectories imply autarky-level displacement only around 2040–2050 (Fig \ref{fig:rollout_trajectory}\textbf{c}).
For industry, electrification and biomass uptake would need to reach around 75$\,$TWh/a (equivalent to around 7.5$\,$bcm of gas) for a 2035 \textit{Autarky} target (Fig \ref{fig:rollout_trajectory}\textbf{d}).
NECP data from 2024 indicate around 150 TWh of gas displacement (around 15$\,$bcm) by 2030, largely from France and Spain, though EU-level policy does not foresee substantial industry electrification before 2030 \cite{EuropeanCommission2022REPowerEU, AgoraIndustry2026ElectrifyingHeat}.


There is some interchangeability between the sectors that deliver these targets.
If industry heat fails to phase out gas, the shortfall would likely be covered by more expensive options, such as building heat.
We find that, without industry electrification, reaching 200$\,$bcm of gas consumption (\textit{Autarky}) becomes about \euro10\,bn/a more expensive and requires roughly 5--6 million additional homes to be equipped with electric heating, assuming biomass can contribute a further 400\,TWh of industry heat (Figure \ref{fig:slow_industry_electrification_progress}).

%% file: sections/results3.tex
\begin{figure}[h]
    \centering
    \includegraphics[width=1\linewidth]{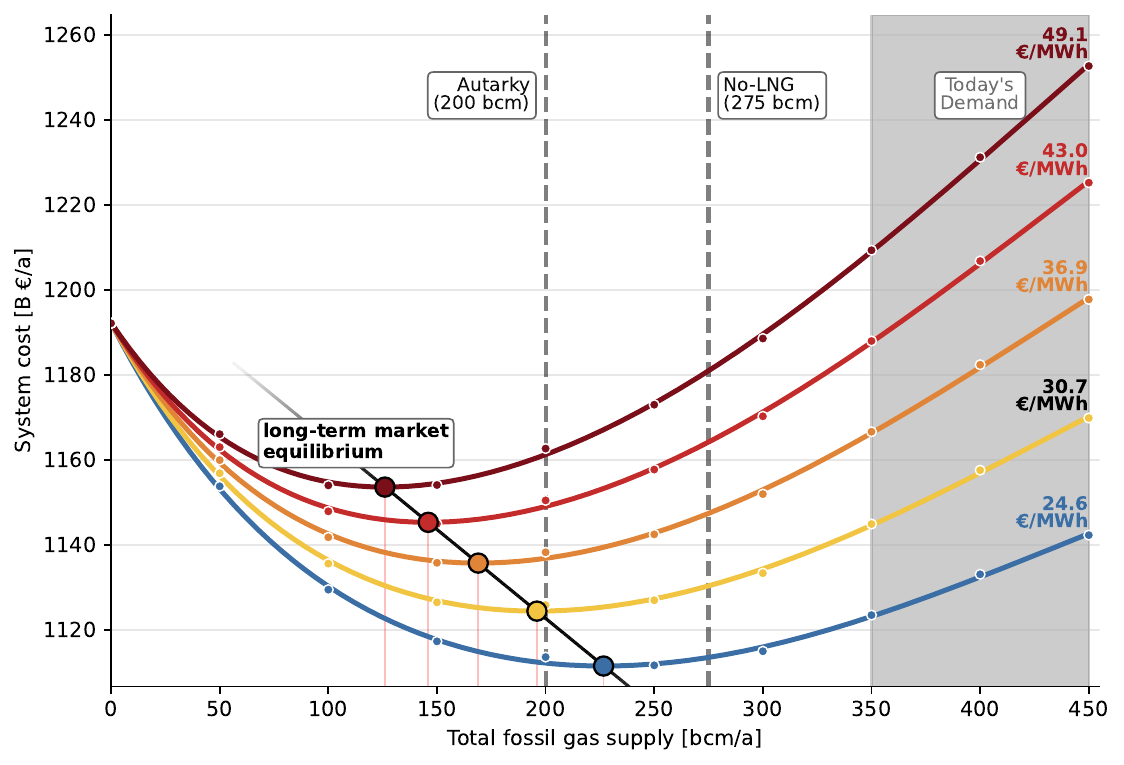}
    \caption{
      System cost under different fossil gas supply constraints and global gas market prices.
      These prices are exogenous to the system and do not include a scarcity rent, unlike in Fig \ref{fig:results_1_figure}\textbf{b}.
      The gas price range of 24.6--49.1$\,$\euro/MWh represents where LNG market prices may settle in the long term.
      Model runs are in steps of 50$\,$bcm (see small markers), and the long-term market equilibrium is derived from quadratic interpolations between them.
      The minimum of each parabola refers to the level of gas consumption where the value of gas to the system coincides with its price (i.e. the scarcity rent is zero), interpretable as the consumption where the model would settle under idealised market conditions.
      The total system cost includes investment, and therefore points on the parabolas to the right of the optimum represent costs for an energy system that is slow to adapt to changes in the price signal.
      Throughout, the model assumes a carbon price of 100$\,$\euro/tCO2.
  }
    \label{fig:long-term-equilibria-gas-use}
\end{figure}

It remains unclear where global gas prices will settle in the long term, but they are unlikely to return to typical pre-2020 or indeed 20th century levels below 25$\,$\euro/MWh \cite{tradingeconomics_ttf_2026}---when the vast majority of existing gas infrastructure was built \cite{acer_mmr_gas_2020}.
Energy infrastructure has lifetimes of decades, so the system is too inert to keep up with changes in commodity prices, a dynamic similar to ``emission lock-in'' \cite{tong2019committed}.
As a result, when gas prices spike, the incumbent capacity build-out commits the energy system to more gas use than would emerge in a long-term equilibrium.
This effect is likely to be compounded as the second Emissions Trading Scheme (ETS2) and the Carbon Border Adjustment Mechanism (CBAM) \cite{zwickl2026eu} expands carbon pricing to additional end-use sectors currently outside the EU ETS, in particular buildings and road transport \cite{euets_directive_2023_959}.

Setting a fossil gas supply constraint, as we do throughout this paper, lets us explore systems that deviate from the long-term economically optimal gas consumption (Fig \ref{fig:long-term-equilibria-gas-use}), replicating a system unable to adjust its long-term infrastructure investment to short-term price changes.
Tracking system costs while varying gas consumption yields a parabola whose minimum defines the long-term cost-optimal gas consumption (long-term because investment is enabled).
In other words, for a given gas price, this is where consumption would settle if it were unconstrained -- where the value of gas to the system equals its price (by duality in linear programming \cite{boyd2004convex}).

We find that if long-term gas prices settle at 30$\,$\euro/MWh or above, near-term carbon pricing (100$\,$\euro/tCO2) drives the system to autarkic gas-supply levels (Fig \ref{fig:long-term-equilibria-gas-use}).
Moreover, gas consumption will only remain above 275$\,$bcm, necessitating LNG imports, if gas prices return to pre-2020 levels of around 15--25$\,$\euro/MWh \cite{acer_mmr_gas_2020}.

We find that, accounting for the carbon price, the long-term elasticity of gas consumption to the external gas price is around $-1.33$ (Methods \textit{Elasticity}, Fig \ref{fig:long-term-equilibria-gas-use}).
This is a relatively high value compared with the long-term elasticities reported in the literature (\cite{Labandeira2017} reports $-0.57$).
However, the frictions captured in empirical work such as \cite{Labandeira2017} (for example capital-stock inertia or imperfect information) are omitted in the linear-optimisation modelling used here; a higher value is therefore expected.

%% file: sections/results2.tex
\begin{figure}[t]
    \centering
    \includegraphics[width=.49\textwidth]{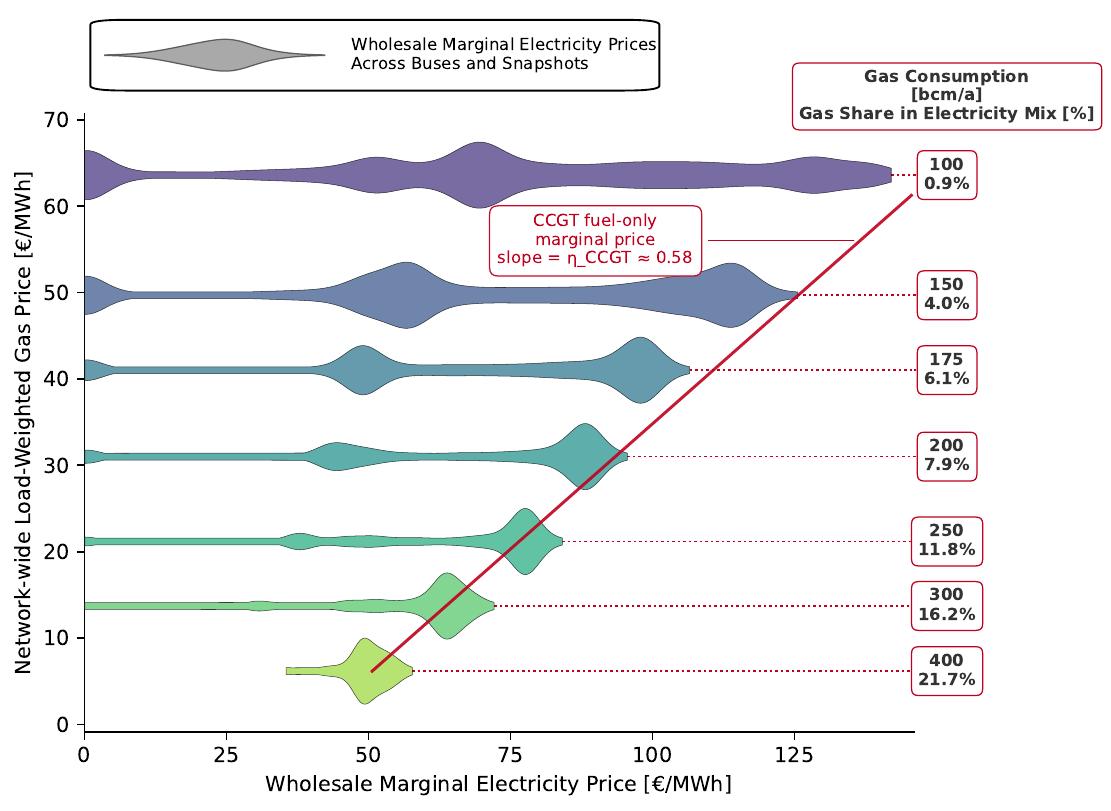}
    \caption{
      Distribution of electricity marginal prices across networks under different fossil gas supply constraints.
      Each violin shows the distribution of electricity marginal prices at low-voltage (consumer) buses, with every bus--snapshot pair contributing one sample.
      Its $y$-position is given by the network-wide load-weighted gas price, which rises with scarcity as consumption decreases; the right-hand labels give the corresponding gas consumption and gas share of the electricity mix.
      The red line shows the fuel-only short-run marginal cost of combined-cycle gas turbines (CCGT), with slope set by their efficiency ($\eta_\mathrm{CCGT} \approx 0.58$).
      Each network has around 146,000 bus--snapshot pairs.
    }
    \label{fig:price_setting}
\end{figure}

A long-standing concern is that, even when gas accounts for only a small share of the power mix, its price can exert a disproportionate effect on consumer electricity prices.
The first mechanism driving this is marginal pricing: in Europe's uniform-price electricity auctions, the marginal generator sets the clearing price for most dispatched power \cite{zakeri_role_2023, schittekatte2023power}.
The second is the anchoring of battery dispatch prices to the opportunity cost of dispatching when gas sets prices \cite{brown2025price, geis2026managing}.
Therefore, phasing out gas power may still fail to shield consumers from that price.
This is directly relevant here: the gas scarcity we simulate drives the model's gas shadow price above 100\,\euro/MWh (see Methods \textit{Fossil Gas Supply Constraint and Gas Price}).

That battery dispatch prices could be driven by the gas price might seem counter-intuitive;
batteries and demand flexibility are generally expected to stabilise electricity prices between near-zero levels --- when renewables saturate the market --- and price spikes in the small subset of hours when dispatchable generators have to recover their investment costs \cite{brown2025price}.
When batteries engage in such arbitrage (smoothing intermittent renewable supply, and, crucially, preventing backup generation by thermal plants), the question is at what price they should discharge.
In the logic of capacity-expansion models (such as PyPSA-Eur), that price is set by the cost of what batteries displace.
When batteries prevent a gas turbine from dispatching, their effective asking price is the short-run marginal cost (SRMC) of the displaced gas turbine.
Thus, even as batteries push fossil generation out of the system, fossil-fuel prices continue to influence electricity prices via battery opportunity costs.

We find that thermal-power marginal costs are the dominant driver of electricity marginal prices.
This emerges from the distribution of these prices across all network regions and snapshots (Figure \ref{fig:price_setting}).
Even when gas contributes less than 8\% of electricity, we find the marginal price anchored by gas around 50\% of the time.
An upper structure in the distribution follows the short-run marginal cost of combined-cycle gas turbines (CCGTs): it increases roughly linearly with the gas price, with a slope set by the CCGT efficiency ($\eta_\mathrm{CCGT} \approx 0.58$; red line).
At even tighter fossil gas supply constraints, the distribution broadens as biomass CHPs become the predominant dispatchable power source in the model, and the resulting coupling to heating-sector marginal prices begins to obscure this relationship.

%% file: sections/results5.tex
We assess system- and consumer-cost increase that would occur during a gas-price hike of the kind driven by contractions in global gas supply, as in 2022 or 2026.
Taking the investment-optimised model with fossil gas supply 400$\,$bcm as baseline, we run an \enquote{operation-only} case in which the model cannot change installed capacities but re-optimises dispatch decisions, reflecting the limited responsiveness of a system facing a short-term gas-price hike.

\begin{figure}[t]
    \centering
    \includegraphics[width=\linewidth]{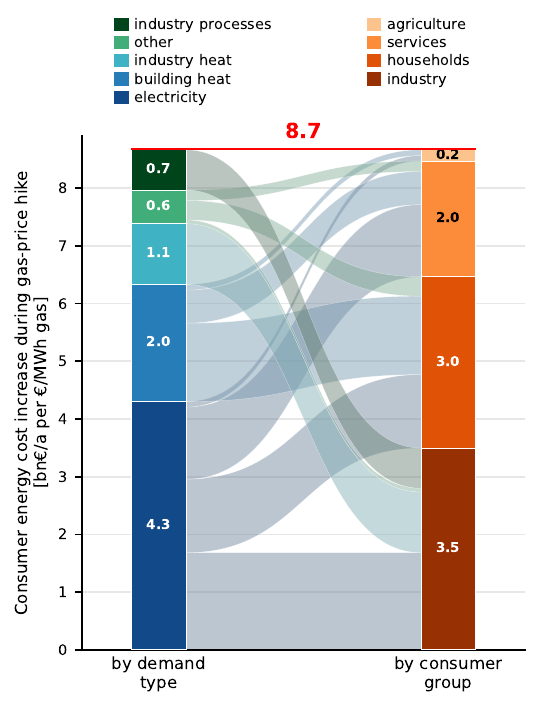}
    \caption{
      Consumer cost markup during a gas-price hike, i.e.,\ an increase in the exogenous gas price (interpretable as a rise in global gas markets), disaggregated by sector and consumer group.
      Results shown for a system at 400$\,$bcm/a gas consumption.
      See Methods for details.
    }
    \label{fig:gas_price_hike_impact}
\end{figure}

We find that a year-round gas-price increase of 1$\,$\euro/MWh raises consumer costs by $\sim$8.7$\,$bn\euro\ (Fig \ref{fig:gas_price_hike_impact}; see Methods, \enquote{Gas Price Hike}).
This is roughly double what the product of gas volume and price increase would suggest (400$\,$bcm, i.e.\ 4,000$\,$TWh, $\times\,$1$\,$\euro/MWh $\approx$ 4$\,$bn\euro), because electricity marginal prices pass the higher gas cost through to other generation technologies.

%% file: sections/discussion.tex
Geopolitical uncertainty has highlighted the risk inherent in Europe's current level of gas imports and motivated policymakers to seek options for curbing the continent's import dependence, for instance by terminating all imports of Russian fossil fuels \cite{EC2025RussianGasAgreement}.
Eliminating LNG imports (i.e.\ lowering consumption to 275 bcm, at which demand can be met by domestic production and pipeline imports) raises annual system cost by $\sim$5 bn\euro ($<$1\%).
Aligning consumption with European production capacity alone ($\sim$200 bcm, autarky scenario) raises it by $\sim$16 bn\euro ($<$2\%).


Electrification of low-temperature industry heat appears to be the most efficient route for displacing around 50$\,$bcm of gas use.
If industry heat fails to deliver its share of gas phase-out and the shortfall is covered by building heat and increased industry biomass heating, autarkic system costs are likely to increase by \euro4--10\,bn (Figure \ref{fig:slow_industry_electrification_progress}).
The central role of industry heat is driven by high utilisation, which is typical of industry use and eases the burden of heat pumps' high investment cost.
Indeed, the techno-economics underlying the model already appear to capture this advantage.
For instance, if the (retail) electricity price is double the gas price, and the carbon price is 100$\,$\euro/tCO2, heat pumps deliver heat at a Levelised Cost of Heat (LCOH) roughly 10--20$\,$\euro/MWh$_\text{th}$ lower than a gas boiler (Fig \ref{fig:heat_pump_gas_boiler_breakeven}), a finding that aligns with previous research that, under the same power-gas price ratio, locates the break-even between gas boilers and heat pumps at 40$\,$\euro/tCO2 \cite{AgoraIndustry2026ElectrifyingHeat}.

The model's findings reflect this fundamental trade-off from its social planner's perspective.
However, translating them to real-world investment decisions hinges on real-world retail price ratios reflecting those in the model, in particular for electricity versus gas.
This is not guaranteed.
On top of the wholesale price (mostly captured in the model), real retail prices add grid fees (partially captured) and taxes (not included), which can muddle the picture.
This holds for both electricity and gas.
Nonetheless, related work suggests the model's findings are coherent with the picture emerging in 2025 European Union energy markets \cite{iea2026electrification} (with increasing 2030 carbon levies poised to tilt the market further toward heat pumps).
The partial perspective of the model yields electricity-to-gas price ratios similar to those assessed in \citet{iea2026electrification}, which concentrate between 2 and 3.2, and identifies large cost-saving electrification opportunities, similar to those we find here.

Findings on residential heat require the same qualification with regard to retail energy prices, and their inaccuracy is compounded in the household context by inter-country differences in heat-pump overnight costs; in CAPEX terms, heat pumps in Germany are twice as expensive as in the UK (Fig \ref{fig:heat_capex_opex_lcohs}).
The present-day assessment of \citet{iea2026electrification}, however, comes to the same outcome as our study: residential heat pumps roughly break even with gas boilers with some regional differences (Fig \ref{fig:heat_pump_gas_boiler_breakeven}).
This is because the model captures several of the factors that drive this heterogeneity.
For instance, the model's temporal resolution captures cold-spell-driven capacity needs both of the heating technology and of the grid.

Distribution-grid expansion, a necessary condition for electrification, conceptually follows a logic similar to that of heat pumps themselves.
Its high-CAPEX, low-OPEX cost structure makes it a smaller economic bottleneck for non-seasonal, high-load-factor industrial demand (Fig \ref{fig:hp_lcoh_disaggregation} and \cite{arlia2026electricity}), but it is also subject to uncertainties that are hard to reflect in linear optimisation modelling.
Uneven network tariffs across European countries \cite{arlia2026electricity, iea2026electrification} reflect that build cost, geographic context and policy structure all play a role in creating an extremely heterogeneous picture \cite{schittekatte2018futureproof, mitei2016utility}.
Overall, both this work and the related literature assess distribution-grid capex to be a smaller cost factor than the investment cost of the heat pumps themselves \cite{iea2026electrification}.

Instead of economics, other factors appear to be the roadblocks to gas phase-out.
For instance, on top of distribution-grid economics are hard barriers to industry electrification stemming from grid-access queue times \cite{ec2025gridconnections}.
Around half of surveyed countries report being unable to integrate further industrial demand \cite{Cremona2026CrossedWires}, and there is wide variation both in the process that manages access and in the priority given to integrating other assets, such as renewable generators \cite{eurelectric2025backlog}.
Industry gas use, where electrification is most economical, is located in the low-temperature band.
Compared with high-temperature demands, these industries are more likely to fall below the 20$\,$MWth heat-demand threshold that would include them in the EU's Emissions Trading Scheme (ETS) \cite{euets_directive_2023_959, dehst2024smallemitters}.
This threshold would effectively be removed in 2028 via the introduction of the ETS2 for most of the remaining sectors \cite{ec2026ets2}, likely leading to a further increase in grid-connection requests.

It is often assumed that similar bottlenecks plague heat-pump uptake in households, in particular that workforce shortages constrain it \cite{iea_future_heat_pumps_2022}.
However, other research contends that the bottleneck lies elsewhere, and indeed leads back to policy failure.
In Germany, for instance, qualified heat-pump installers currently complete an average of three installations per person per year \cite{altermatt2023replacing}.
This is well below the rate this workforce could sustain, given that a heat-pump installation typically requires only 4--6 person-days.
Similar situations have been observed in other countries such as the United Kingdom and the Netherlands \cite{nesta2024installmore, ehpa2024dutch}, highlighting the role of weak demand pull.
A likely driver is inconsistent policy incentives, which forgo the potential for a meaningful acceleration in roll-out and likely manifest as a German heat pump costing twice as much as a British or Portuguese one (Fig \ref{fig:heat_capex_opex_lcohs}).

We further highlight that our findings are conditional on carbon pricing expanding to previously uncovered sectors, especially building heat and small industries.
As found both here and in other work, carbon pricing critically aligns decarbonisation with energy-independence efforts \cite{jewell2016comparison}, and without it fossil-import reduction pulls domestic fossil energy sources such as lignite back into the system \cite{pedersen2022long, Victoria2020}.
This was also observed in the 2022 gas crisis, during which EU year-on-year coal generation rose by 7\%.

The risk consumers face during these crises should frame the discussion of the cost of import independence.
We find that currently, at around 400$\,$bcm gas use, consumer bills rise by $\sim$8.7$\,$bn\euro for every \euro/MWh rise in global gas prices.
This scale matches the markup Europe paid during the 2022 gas crisis: extrapolated to the $\sim$100$\,$\euro/MWh rise in the 2022 year-average gas price relative to 2021, it implies a burden on the order of $\sim$710$\,$bn\euro, close to the 300--400$\,$bn\euro\ import markup \cite{iea_gas_2022_drivers} and the $\sim$700$\,$bn\euro\ of consumer support deployed that year.
We caution, however, that the model likely overestimates flexibility to forgo gas in the power sector and underestimates demand elasticity in industry, which in 2022 was the most price-responsive gas consumer \cite{ruhnau2023natural}.
We therefore find that the cost of the \textit{No-LNG} scenario is lower than the consumer energy burden of a sustained 1$\,$\euro/MWh rise in gas prices.

Finally, we highlight the need for policy support that ensures lower gas consumption translates into lower consumer exposure to global gas prices.
We find that gas has a large effect on marginal electricity prices, even when it supplies only a low-single-digit share of the power mix.
Marginal pricing is what causes the aforementioned superlinear relationship between external gas prices and consumer cost.
This dynamic will become more pronounced in a future system, in which electricity increasingly becomes the backbone of energy supply.
Taken together, these dynamics suggest that decoupling bulk (renewable) electricity generation from marginal prices (for instance through two-sided CfDs) will likely be crucial to protect consumers from global gas shocks, even when gas use has been greatly curbed.

In conclusion, this work shows that Europe's path away from gas imports is not only strategically attractive, but also economically compelling:
The cost of achieving import independence is substantially lower than the risk premium Europe has paid on those imports since 2022.
Moreover, the policy framework already in place, combined with gas prices prevailing since 2020, generates enough market incentive to drive gas consumption down to levels consistent with domestic production capacity by 2030--35.
The remaining challenge for policy is to dismantle the regulatory, infrastructural, and informational barriers that prevent the market from acting on the cost signal it already receives, and to ensure that curbing gas imports actually shields energy consumers from their exposure to swings in gas prices.

%% file: sections/methods.tex
\subsection*{PyPSA-Eur}

This work uses the PyPSA-Eur model \cite{neumann2023potential}, an open-source, sector-coupled, linear energy-system model of Europe that co-optimises investment and operation.
We run it at a 3-hourly temporal resolution with a spatial resolution of 50 nodes, covering the power, heat, biomass, industry, transport and agriculture sectors.
The study performs an overnight optimisation of a brownfield system that includes the capacities existing in 2024 and uses 2024 weather data.
No hard limit is placed on carbon emissions; instead, a carbon price of 100\,\euro/tCO2 is applied.
The model can expand transmission capacity by up to an additional 25\% of current installations.
We use the standard version of the model, with the additional disaggregation of industry heat demand into distinct temperature bands described below.
Our assumptions on biomass and biogas availability are largely based on the JRC ENSPRESO dataset, from which we use the \textit{Medium Potential} scenario \cite{ruiz_enspreso_2019, millinger2025diversity}.

The model is based on the PyPSA modelling framework \cite{brown2017pypsa}.
Its fundamental logic is a nodal balance, at each node and timestep, between the energy supplied and the combined withdrawals.
On top of this logic sit energy demands, along with generation, conversion, storage and energy transport (such as transmission lines or pipelines).
Each has capital and marginal costs and technical constraints, such as spatial limits on renewable build-out.
When the optimisation problem is solved, a marginal price of an energy carrier emerges at each node and timestep, the price of producing one more unit of that carrier at that node and timestep.
This price is a dual variable: the change in the optimisation's objective function if another unit were produced for free.
Therefore, it is a close proxy for real marginal price formation.

\subsection*{Fossil Gas Supply Constraint and Gas Price}

The main results of this paper derive from forcing the model to use specified volumes of fossil gas: the constraint is an equality, not an upper bound, so the model uses exactly the prescribed volume.
Gas has a uniform price of 24.6 \euro/MWh, regardless of its source (except where different gas prices are explored explicitly, Fig \ref{fig:long-term-equilibria-gas-use}).
The PyPSA-Eur model works in MWh, i.e. TWh for larger quantities.
For the exposition we however chose to show natural gas volumes in billion cubic meters (bcm) instead of TWh, using a simple conversion factor of 10, given that conventation vary around this value.

On top of the exogenous price, we observe gas prices that emerge endogenously during optimisation as shadow prices, and report these in our findings.
Though phenomenologically similar, price formation in the model is not a simulation of scarcity pricing in a real market.
Instead, it is the dual variable of the fossil gas supply constraint, quantifying the change in the objective function were the constraint relaxed by one MWh.
It is the price level required to elicit the corresponding level of fossil gas supply.
Similarly, in a real market the willingness to pay for gas would be driven by the cost of the next-best supply option.

\subsection*{Industry Heat Demand}

As a novel feature, we represent industry heat demand in four distinct temperature bands, estimating the magnitude of each per region from local industry presence.
This estimation takes regional sectoral industry outputs and multiplies them by the share of demand that falls in each temperature band.
This is because the technologies available to meet a heating demand are determined by the quality of the heat.
For instance, above 500\deg C only methane and hydrogen are generally applicable.
Up to almost 200\deg C, industrial heat pumps are a viable option.

In detail, we start from the JRC IDEES dataset to obtain country-level industry production outputs in tonnes, covering the sectors shown in Fig \ref{fig:industry_sectors_production}.
The IDEES dataset also includes input energy-carrier volumes for industry production, and further includes \textit{best-in-class} processes---the production pathways most likely to provide a carbon-neutral variant.
For example, in the steel sector, most current production uses the coke-based basic oxygen furnace (BOF) route.
The carbon-neutral variant is hydrogen direct reduction (H2-DRI) paired with an electric arc furnace (EAF), which is emissions-free when green hydrogen is used.
Past versions of PyPSA-Eur make exogenous assumptions about the roll-out of these technologies, e.g.\ assuming that by 2035 half of steel is produced by H2-DRI.
These assumptions are then translated into primary energy demands in the model, in this case mostly coke for BOF, and hydrogen and electricity for the H2-DRI route.
The model then only optimises the supply of these input commodities.

For heating demands, which make up a large share of the total, the input carrier providing the heat is interchangeable, provided it can deliver heat at the required temperature and does not also serve as a feedstock (as for instance coke does for steel).
As a result, we treat the large majority of industrial heat currently supplied by solid biomass, gas and waste heat as generic, except in steel and some chemical production.
Fig \ref{fig:industry_temp_band_shares} shows the share of heating demand per temperature band for each industry product considered, which we multiply by the total heat input per tonne of industry product.
The total heat input is estimated from JRC IDEES \cite{jrc_idees_2021} as the total MWh value of solid biomass, gas and waste heat.

Since this work focuses on gas consumption, we endogenise only the heat demand currently met by gas, and assume the remainder continues to be met by its original carrier.
We cannot know in which temperature band gas heating is deployed; given its ability to deliver high-quality heat, we assume that, when heat demand is sorted by temperature, the hottest X\% is met by gas (Fig \ref{fig:industry_hot_end}), since among the carriers currently in use it is the only one that can generically supply high-temperature heat.

This results in around 1,300 TWh of heat demand currently met by gas (equivalent to around 130$\,$bcm), of which roughly half lies below 500\deg C, with the remainder in high-temperature or feedstock uses (see Fig \ref{fig:industry_heat_demand_per_band}).
Fig \ref{fig:country_industry_heat_demand} shows how this sub-500\deg C demand is distributed across European countries.
For comparison, Agora estimates total European industry heat demand at around 1,860\,TWh \cite{AgoraIndustry2026ElectrifyingHeat}.
Both in aggregate and per sector, our numbers are broadly consistent with this.

We offer the model a range of technologies to meet heat demand in each temperature band (Tab \ref{tab:industryheat}).

\subsection*{Building Heat Demand}
\label{subsec:building_heat}

Residential and service-sector heat demand is split in each region into rural, individual urban, and district urban heating.
The split between residential and service-sector demand follows shares derived from population-weighted national energy balances (roughly 70\% residential, 30\% services across Europe); within each settlement category the two share a common set of heating technologies but are tracked separately, letting us attribute consumer costs to either group (see the gas-price hike analysis below).
Each demand is met by a mix of technologies, typically gas boilers, biomass boilers, heat pumps and solar thermal (see \citet{zeyen2021mitigating} for a detailed discussion).
See Table \ref{tab:buildingheat} for an overview of costs.
For heat pumps (air and ground) the efficiency represents the coefficient of performance, a quantity that varies by time and region.

The model is brownfield and therefore includes existing heating capacities.
Country aggregates of these are shown in Fig \ref{fig:existing_heating_capacities}.

However, without additional constraints, the model risks overestimating consumers' supply flexibility.
Unlike electricity, for instance, urban individual and rural heat are not traded in liquid markets in which consumers can make operational decisions about their supply.
Instead, consumers are committed to the heating technology they have installed (i.e.\ to their investment decision).
The model is unaware of this and instead tends to operate a pooled market, in which low-OPEX technologies provide baseload heat (typically heat pumps) and low-CAPEX technologies meet demand peaks (typically boilers).

\begin{figure*}[t]
    \centering
    \includegraphics[width=0.8\textwidth]{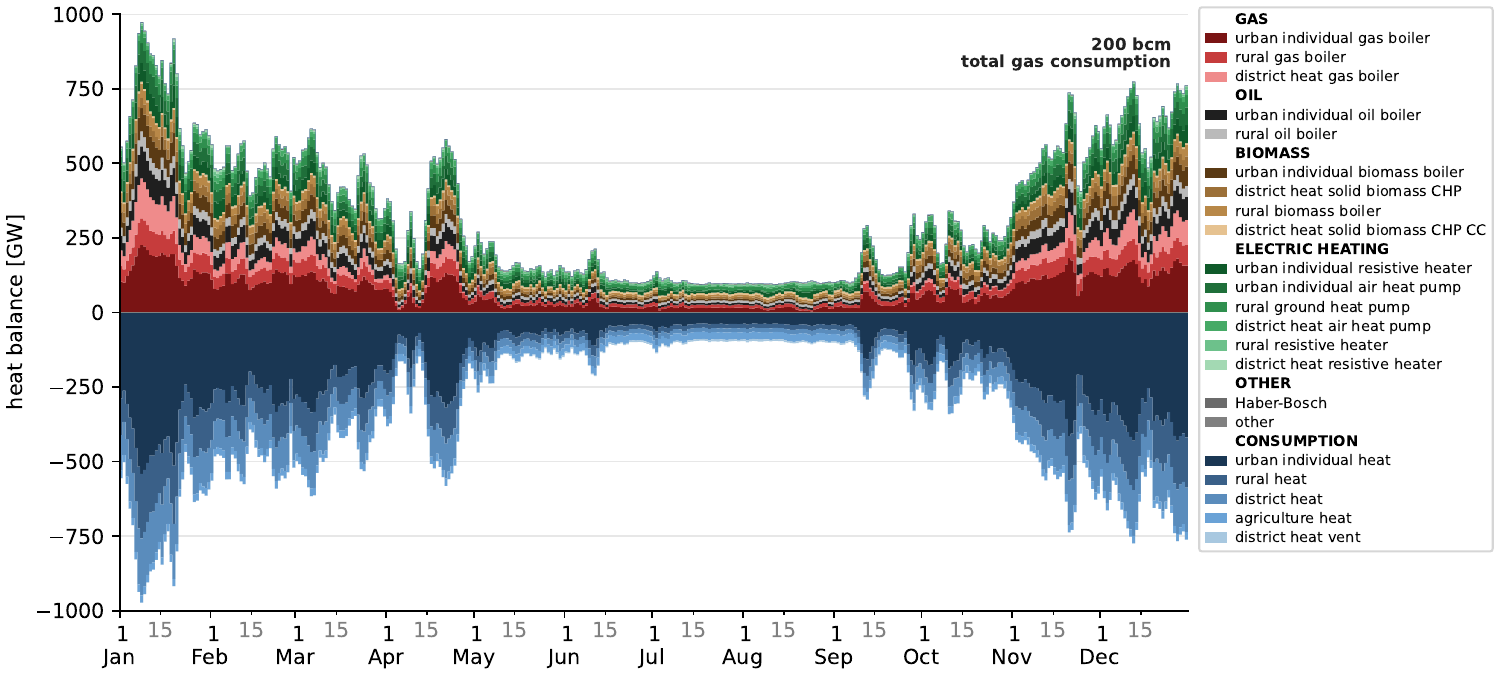}
    \caption{
      Aggregate heating supply and demand for a model with 200\,bcm gas consumption.
    }
    \label{fig:gas_timeseries}
\end{figure*}

To address this, we constrain heating technologies in urban individual and rural heat to be \textit{load-following}, i.e.\ their partial load during time $t$ is bounded above by the relative load $l_t/l_{\text{max}}$, where $l$ is that bus' time series of heat load and $l_{\text{max}}$ is its maximum (Fig \ref{fig:gas_timeseries}).
This constrains all technologies to a partial load of around 25\%, and, as we find, yields LCOH values that align well with the literature (Fig \ref{fig:lcohs}).
This method is not applied to district heating, where heat plants can typically modulate, to some extent, between heat-production pathways \cite{lund_4gdh_2014}.

\begin{table}[htbp]
\centering
\caption{Technology options for industrial process heat supply by temperature band (2030 projections).
Investment costs are per unit of input capacity and include the efficiency scaling applied in the model ($c_\text{inv} \times \eta$).
All monetary values in EUR\textsubscript{2020}.}
\label{tab:industryheat}
\footnotesize
\setlength{\tabcolsep}{3pt}
\begin{tabular}{@{}llccrcc@{}}
\toprule
\textbf{Temp.} & \textbf{Technology} & \textbf{Input} & $\boldsymbol{\eta}$ & \textbf{Inv.} & \textbf{VOM} & \textbf{Source} \\
 &  &  &  & {\scriptsize\euro/kW\textsubscript{in}} & {\scriptsize\euro/MWh} & \\
\midrule
\multirow{4}{*}{\rotatebox[origin=c]{90}{\scriptsize$<$\,100\,°C}}
 & Biomass boiler        & Biomass     & 0.89 & 530   & 2.84 & \cite{DEA_IPH} \\
 & Gas boiler            & CH\textsubscript{4} & 0.93 & 43    & 1.01 & \cite{DEA_IPH} \\
 & Heat pump (med.\ T)   & Elec.       & 2.70 & 2\,118 & 3.22 & \cite{DEA_IPH} \\
 & Electric boiler       & Elec.       & 0.99 & 70    & 0.88 & \cite{DEA_IPH} \\
\midrule
\multirow{4}{*}{\rotatebox[origin=c]{90}{\scriptsize 100--200\,°C}}
 & Biomass boiler        & Biomass     & 0.89 & 530   & 2.84 & \cite{DEA_IPH} \\
 & Gas boiler            & CH\textsubscript{4} & 0.93 & 43    & 1.01 & \cite{DEA_IPH} \\
 & Heat pump (high T)    & Elec.       & 3.05 & 2\,870 & 3.22 & \cite{DEA_IPH} \\
 & Electric boiler       & Elec.       & 0.99 & 70    & 0.88 & \cite{DEA_IPH} \\
\midrule
\multirow{3}{*}{\rotatebox[origin=c]{90}{\scriptsize 200--500\,°C}}
 & Direct firing (solid) & Biomass     & 1.00 & 222   & 9.33$^\S$ & \cite{DEA_IPH,Visser2020} \\
 & Direct firing (gas)   & CH\textsubscript{4} & 1.00 & 15    & 0.28 & \cite{DEA_IPH} \\
 & H\textsubscript{2} combustion & H\textsubscript{2} & 1.00 & 151   & 2.79 & own est.$^*$ \\
\midrule
\multirow{2}{*}{\rotatebox[origin=c]{90}{\scriptsize $>$\,500\,°C}}
 & Direct firing (gas)   & CH\textsubscript{4} & 1.00 & 15    & 0.28 & \cite{DEA_IPH} \\
 & H\textsubscript{2} combustion & H\textsubscript{2} & 1.00 & 151   & 2.79 & own est.$^*$ \\
\bottomrule
\end{tabular}

\vspace{4pt}
\raggedright\scriptsize
All data compiled via \cite{PyPSA_techdata}. \quad
$^\dag$\,Incl.\ post-combustion capture unit at 2.7\,M\,\euro/(tCO\textsubscript{2}/h); 90\,\% capture rate. \quad
$^\ddag$\,$\eta$ reduced by CCS heat demand. \\
$^\S$\,Incl.\ pelletising cost of 9\,\euro/MWh. \quad
$^*$\,Assumed $10\times$ gas direct firing cost (no established reference).
\end{table}

%
%

\begin{table}[htbp]
\centering
\caption{Technology options for building heat supply by heat system (2030 projections).
Investment costs are per unit of input capacity and include the efficiency scaling applied in the model ($c_\text{inv} \times \eta$).
All monetary values in EUR\textsubscript{2020}.}
\label{tab:buildingheat}
\scriptsize
\setlength{\tabcolsep}{2pt}
\renewcommand{\arraystretch}{1.05}
\begin{tabular}{@{}llccrcc@{}}
\toprule
\textbf{System} & \textbf{Technology} & \textbf{Input} & $\boldsymbol{\eta}$ & \textbf{Inv.} & \textbf{VOM} & \textbf{Source} \\
 &  &  &  & {\tiny\euro/kW\textsubscript{in}} & {\tiny\euro/MWh} & \\
\midrule
\multirow{7}{*}{\rotatebox[origin=c]{90}{\tiny Rural}}
 & \shortstack[l]{Air-source\\heat pump}    & Elec.       & 3.60 & 3\,238 & --    & \cite{DEA_heating} \\
 & \shortstack[l]{Ground-source\\heat pump} & Elec.       & 3.90 & 5\,778 & --    & \cite{DEA_heating} \\
 & Resistive heater        & Elec.       & 0.90 & 95     & --    & \cite{PyPSA_techdata} \\
 & Gas boiler              & CH\textsubscript{4} & 0.98 & 308 & -- & \cite{DEA_heating} \\
 & Oil boiler              & Oil         & 0.90 & 149    & --    & \cite{PyPSA_techdata} \\
 & Biomass boiler          & Biomass     & 0.86 & 591    & 9.00$^\dag$ & \cite{DEA_heating,Visser2020} \\
 & Solar thermal           & --          & --   & 286$^\ddag$ & --  & \cite{PyPSA_techdata} \\
\midrule
\multirow{6}{*}{\rotatebox[origin=c]{90}{\tiny Urban indiv.}}
 & \shortstack[l]{Air-source\\heat pump}    & Elec.       & 3.60 & 3\,238 & --    & \cite{DEA_heating} \\
 & Resistive heater        & Elec.       & 0.90 & 95     & --    & \cite{PyPSA_techdata} \\
 & Gas boiler              & CH\textsubscript{4} & 0.98 & 308 & -- & \cite{DEA_heating} \\
 & Oil boiler              & Oil         & 0.90 & 149    & --    & \cite{PyPSA_techdata} \\
 & Biomass boiler          & Biomass     & 0.86 & 591    & 9.00$^\dag$ & \cite{DEA_heating,Visser2020} \\
 & Solar thermal           & --          & --   & 286$^\ddag$ & --  & \cite{PyPSA_techdata} \\
\midrule
\multirow{6}{*}{\rotatebox[origin=c]{90}{\tiny District heat}}
 & \shortstack[l]{Air-source\\heat pump}   & Elec.       & 3.20 & 2\,900 & 2.66  & \cite{DEA_DH} \\
 & Resistive heater        & Elec.       & 0.99 & 63     & 1.06  & \cite{DEA_DH} \\
 & Gas boiler              & CH\textsubscript{4} & 1.04 & 55  & 1.06 & \cite{DEA_DH} \\
 & Gas CHP                 & CH\textsubscript{4} & 0.41 / 0.41$^\S$ & 243 & 4.44 & \cite{DEA_DH} \\
 & \shortstack[l]{Solid biomass\\CHP}       & Biomass     & 0.27 / 0.82$^\S$ & 957 & 4.85 & \cite{DEA_DH} \\
 & Solar thermal           & --          & --   & 148$^\ddag$ & --  & \cite{PyPSA_techdata} \\
\bottomrule
\end{tabular}

\vspace{4pt}
\raggedright\tiny
All data compiled via \cite{PyPSA_techdata}. \quad
$^\dag$\,Wood-pellet supply-chain cost from \cite{Visser2020} applied as marginal cost on the biomass input. \quad
$^\ddag$\,Solar thermal collectors enter the model as generators on the heat bus; investment is given in \euro/m\textsuperscript{2} of collector area, not per kW\textsubscript{in}. \\
$^\S$\,CHP electric / heat efficiency (per unit fuel input).  Investment is per kW of fuel input (i.e.\ $c_\text{inv,el}\!\times\!\eta_\text{el}$); the heat output is co-produced via the back-pressure coefficient $c_b$.
\end{table}

\subsection*{Gas Price Hike}
\label{subsec:gas_price_hike}

To quantify how a short-term increase in the global gas price affects system costs and propagates to consumers, we perform \textit{gas-price hike} runs on top of a pre-solved system (Fig \ref{fig:gas_price_hike_impact}).
Starting from an investment-optimised network -- here a current-day, gas-dependent system at 400$\,$bcm/a fossil gas supply -- we fix all asset capacities to their optimised values and re-solve the operational problem only, raising the marginal cost of every fossil-gas generator by a uniform markup $\Delta p$ (here $+1\,$\euro/MWh).
Crucially, for these runs we relax the fossil-gas consumption constraint, so that dispatch is free to curb gas wherever the price increase makes an alternative cheaper.
We note, however, that demand elasticity and fuel-switching options are only coarsely represented in the model, so this should be read as an approximate estimate.

Consumer cost is computed as each load's consumption multiplied by the nodal marginal (shadow) price of its energy carrier, summed over the year, since nodal marginal prices reflect the prices seen by consumers.
The markup reported in Fig \ref{fig:gas_price_hike_impact} is the change in this expenditure per unit gas-price increase, $(\,\text{expenditure}(\Delta p) - \text{expenditure}(0)\,)/\Delta p$, in bn\euro/a per \euro/MWh.
We evaluate it directly at $\Delta p = 1\,$\euro/MWh, a small perturbation chosen to keep fossil gas on the price-setting margin.
The response is mildly sub-linear: a large shock would eventually push gas off the price-setting margin, after which further increases no longer lift the marginal price; the small perturbation used here stays in the near-linear regime.

In the main text we disaggregate the markup two ways: once by energy sector and once by consumer group.
For the consumer-group split, each sector's expenditure is attributed to households, services, industry or agriculture, in demand shares derived from population-weighted national energy balances (electricity split evenly between households and services, building heat roughly $70{:}30$ between residential and services, while industrial and agricultural carriers are already separated in the model).

\subsection*{Elasticity}
\label{subsec:elasticity}

We use the textbook point definition of the elasticity $\epsilon$ of a quantity $Q$ with respect to a price $P$ \cite{Labandeira2017}:
\begin{equation}
\epsilon = \frac{\ln(Q^\ast/Q)}{\ln(P^\ast/P)} \: ,
\end{equation}
where $(P,Q)$ is the reference operating point and $(P^\ast,Q^\ast)$ the perturbed one.
The definition can also be expressed in discretised form as $\epsilon \approx (\Delta Q/Q)/(\Delta P/P)$.
We evaluate it for gas consumption responding to the external gas price (Fig \ref{fig:long-term-equilibria-gas-use}).
We consider only this demand-side response; the price elasticity of supply --- by which lower consumption would itself depress the gas price --- is not included (see Limitations).

\subsection*{Limitations}
\label{subsec:limitations}

The study uses a brownfield model, in which existing capacities in the power sector and building heat are already built.
Since existing assets still have to be paid off during their lifetimes, this introduces a slight bias towards using existing assets, reflecting real-life incentives.
However, since existing assets like gas boilers tend to be CAPEX-light, the model still has a strong incentive to switch to new technologies with lower overall costs.
A faster boiler phase-out entails more expensive stranding.
Capturing this dynamic in an overnight model would require embedding the cost of asset stranding into the capital costs of the displacing technology.
If gas boilers have a lifetime of 25 years, the model could split the replacing technologies into 25 slices, gradually cheapened by the share of boilers stranded with $x$ years of lifetime remaining.
This is beyond the scope of the present study, but could be investigated in future research.

In industry, real heat-pump prices vary by application size; we use a single cost value that broadly reflects this range rather than varying it by size.
To examine this, we plot the competitive advantage of heat pumps over gas boilers as a function of demand-profile shape and the ratio of electricity to gas prices (Fig \ref{fig:heat_pump_gas_boiler_breakeven}).
As expected, heat demands that are unevenly spread, i.e.\ have a low load factor (as arises from seasonality or cold spells), work against heat-pump economics.
This is because the heating asset is sized by peak rather than average demand, and heat pumps, which are high-CAPEX and low-OPEX, are less suited to this configuration.
Placing the model's commodity prices and demand profiles on this map, we see that in industry heat pumps hold a firm advantage of 10--20$\,$\euro/MWh\th, while in residential heat they roughly break even with gas boilers.

The study assesses a broad range of gas consumption levels, set within a background system that aims to reflect 2030--35.
We chose to include extremes in the range of gas consumption levels assessed, spanning from zero fossil gas consumption up to mid-2020s levels of 350--450$\,$bcm/a.
These extremes are implausible outcomes for the 2030s, lying far outside the expected range \cite{EuropeanCommission2022REPowerEU}.
They further complicate validation against heat supply (Fig \ref{fig:heat_validation}), as it is likely the higher electricity availability in the 2030s that causes the model to overestimate electric heat supply, even given the inclusion of the existing heating fleet.
In spite of these drawbacks, we believe the value of isolating the impact of gas consumption justifies forgoing a framing in which gas phase-out co-evolves with the wider system, which would introduce substantial noise into the analysis.

The study relies on a single weather year for 2024 to capture recent climate conditions.
However, this misses inter-annual weather variation that could change renewable generation, demand and heat-pump capacity factors, and thereby affect the economics of the building-heat and power sectors.
We find that the COPs and capacity factors, on average, represent a typical weather year (Fig \ref{fig:renewable_cf_maps}).
That said, there was a cold spell during January 2024.
Demand spikes are more burdensome for heat-pump economics, both because heat pumps are CAPEX-intensive and because they require expansion of the distribution grid.
This could introduce a bias against gas phase-out in building heat.
On balance, however, we believe this to be a small effect, as it is mostly localised to northern regions, which host only a small minority of demand; moreover, geographically driven hydropower availability is likely to offset the unusually high demand, as indicated by the low northern electricity prices (see Fig \ref{fig:regional_marginal_prices}).

Real consumer decisions are based on costs that cannot be fully represented within the current logic of the PyPSA-Eur model.
This plays a notable role in the case of electrification, where the cost balance between gas boilers and electric alternatives could be skewed.
For one, the wholesale electricity price---best represented in the model---makes up only a third of the consumer price.
Investment in the transmission and distribution grids is factored into the model's heat-pump economics, and could be added here ex-post.
However, through its annuitised formulation, it is constrained to represent grid cost as a markup of only the network additions needed and delivered within that year.
This does not align with the per-kWh grid tariffs faced by some consumers.
For consumers, grid tariffs remunerate both the sunk costs of past investment and front-loaded investment in the future grid.
A further share of consumer electricity costs consists of utility margins and taxes, neither of which is represented here; overall, however, the macroscopic cost landscape aligns with the literature (see Discussion).

In industry heating, the endogenisation of heat supply disaggregates demand by temperature alone, abstracting from other process characteristics that can bear on the choice of low-carbon alternative.
For instance, many processes below 500$\,$°C are supplied via steam, for which an electric boiler or high-temperature heat pump is a natural substitute.
Kilns, by contrast, transfer heat to the product directly rather than through steam, so their alternatives must accommodate this mode of heat delivery.
For some products, neglecting these distinctions could misrepresent the true cost of alternatives.
However, the process temperature remains the dominant determinant of the cost structure of both incumbent supply and its alternatives \cite{madeddu2020}.
Future work could disaggregate industry heat demand further and, on a process-by-process basis, offer the model more tailored alternatives.

%% file: sections/appendix.tex
\begin{figure*}[t]
    \centering
    \includegraphics[width=\textwidth]{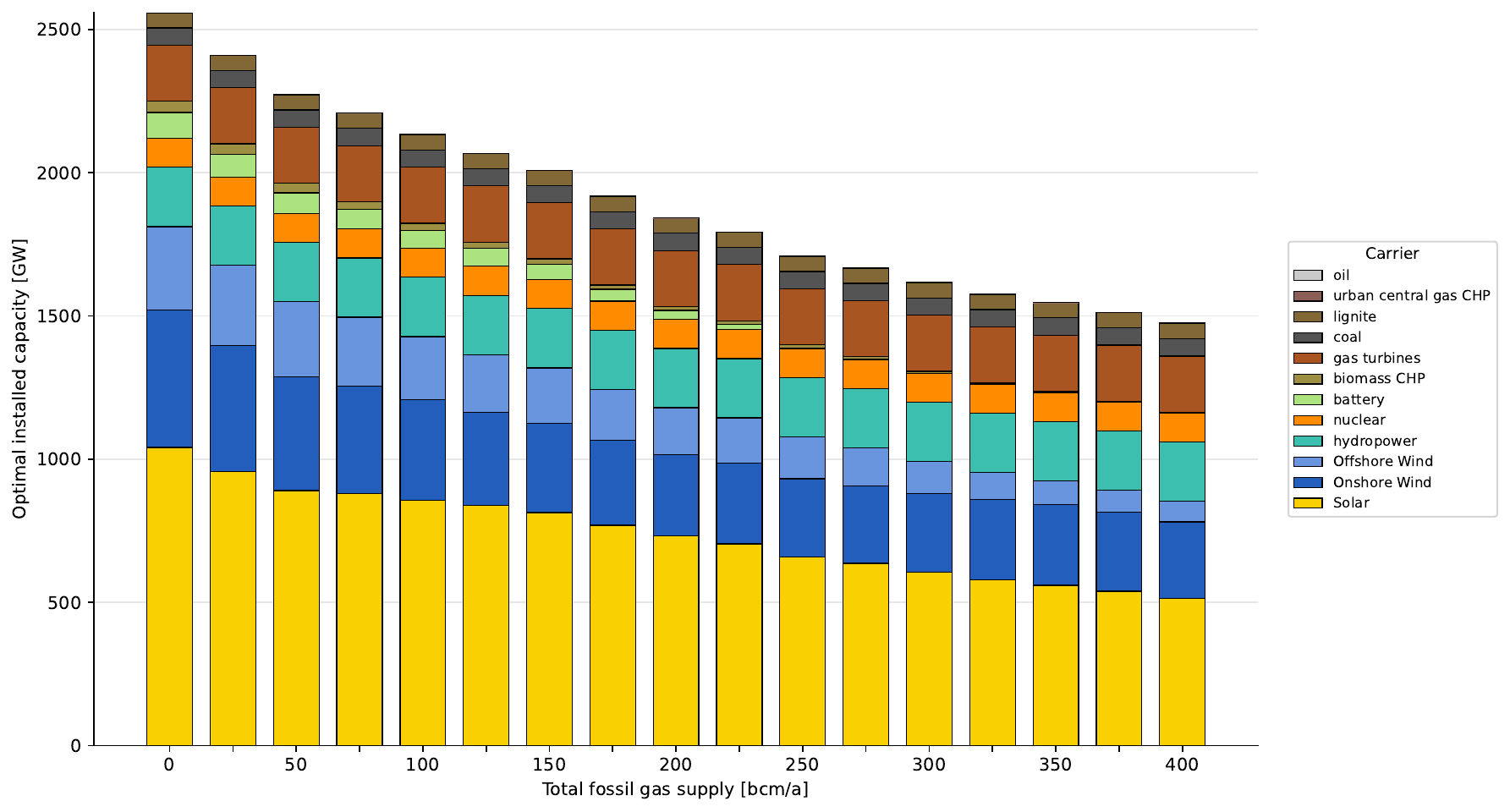}
    \caption{
      Cost-optimal nameplate capacities in the power sector for different levels of gas consumption.
    }
    \label{fig:optimal_power_capacity}
\end{figure*}

\begin{figure*}[t]
    \centering
    \includegraphics[width=0.8\textwidth]{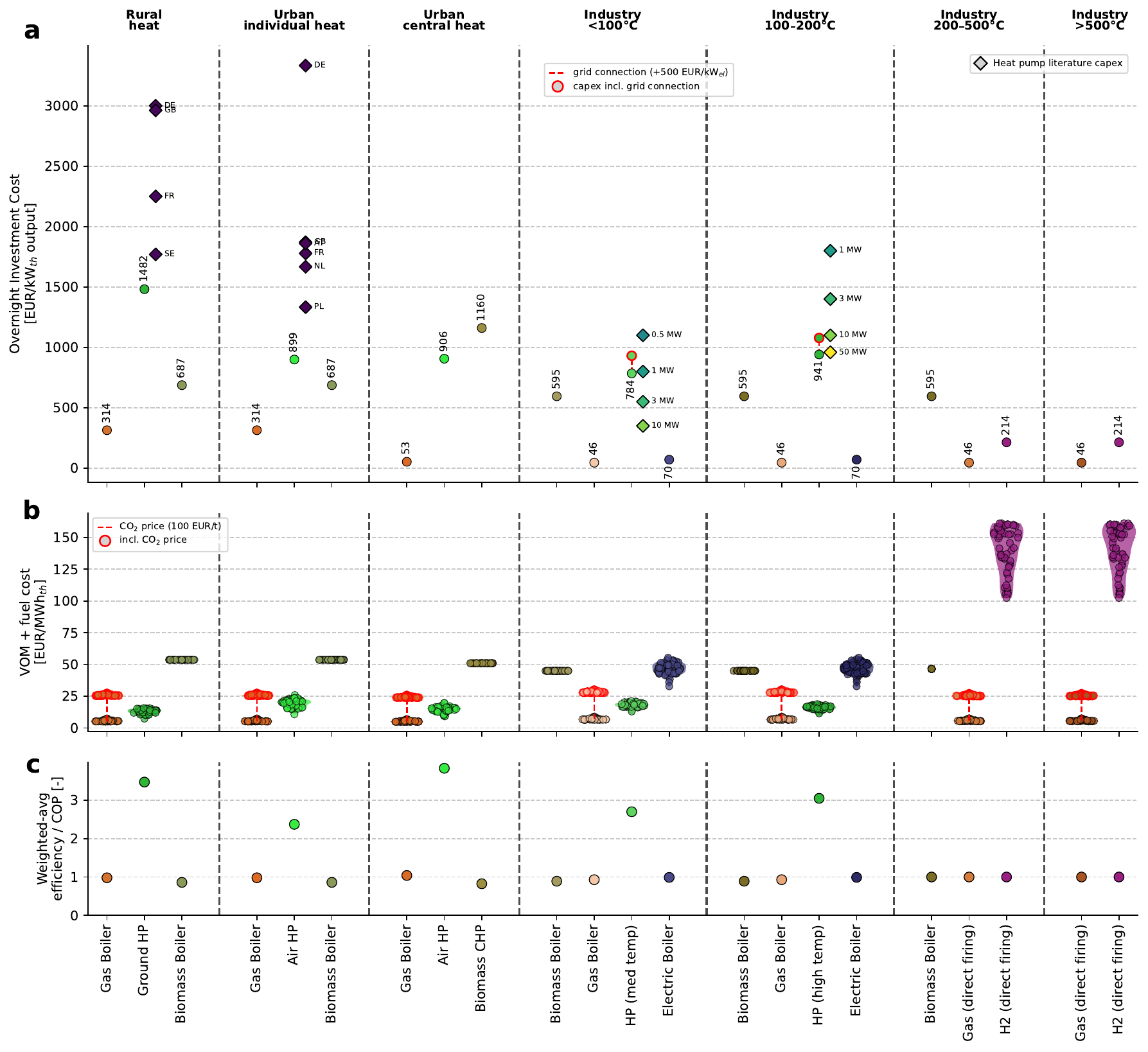}
    \caption{
      Comparison between model values and literature values of \textbf{a} Overnight investment cost, \textbf{b} variable operational plus fuel cost and \textbf{c} efficiency/COP.
      As in the rest of the model, cost assumptions are for 2030, and here are compared with other literature values for 2026.
      Hydrogen fuel costs assume electrolytic production.
      Fuel costs are endogenously estimated from an energy-system model with annual gas consumption of 400$\,$bcm/a.
      Note the values in \textbf{b} are in thermal MWh, and therefore already account for efficiency. Multiplication of the shown values with the respective efficiency/COP approximately yields the fuel cost.
      Literature cost values are compiled from \cite{heatpumpswatch2026, ukbus2025, marina2021industrial, pieper2018allocation, iea2023heatpumpcost}.
    }
    \label{fig:heat_capex_opex_lcohs}
\end{figure*}

\begin{figure*}[t]
    \centering
    \includegraphics[width=0.8\textwidth]{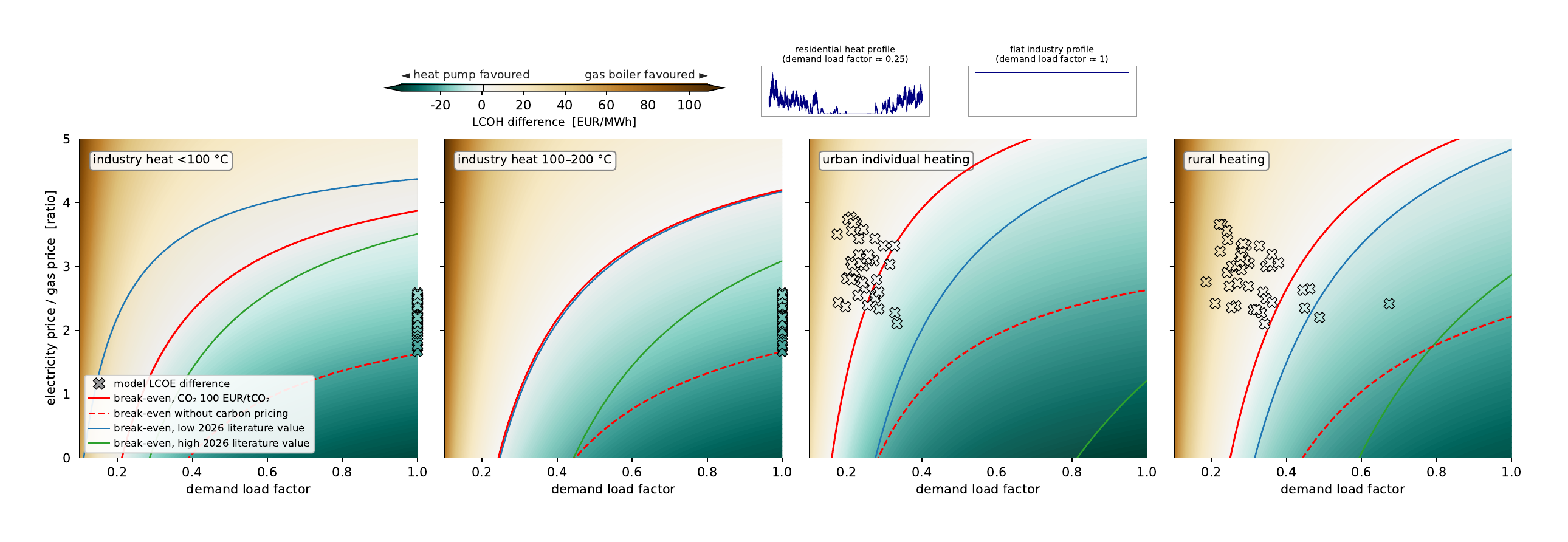}
    \caption{
      Comparison of gas boiler and heat pump Levelised Cost of Heat (LCOH) for different heat demands.
      From the left to right these heat pump technologies are medium-temperature industry heat pump, high-temperature industry heat pump, (residential) air-source heat pumps and ground-source heat pump.
      Each cross refers to the heat pump vs gas boiler trade-off in one region of the network.
      The model LCOHs are taken from a system with gas consumption of 200$\,$bcm/a.
      \textit{Demand load factor} quantifies the variation within a demand profile by dividing the average by the maximum.
      \textit{Break-even} refers to LCOH-parity between heat pumps and gas boilers.
      The high and low literature values are the same as for Fig \ref{fig:heat_capex_opex_lcohs} \cite{heatpumpswatch2026, ukbus2025, marina2021industrial, pieper2018allocation, iea2023heatpumpcost}.
    }
    \label{fig:heat_pump_gas_boiler_breakeven}
\end{figure*}

\begin{figure}[t]
    \centering
    \includegraphics[width=0.49\textwidth]{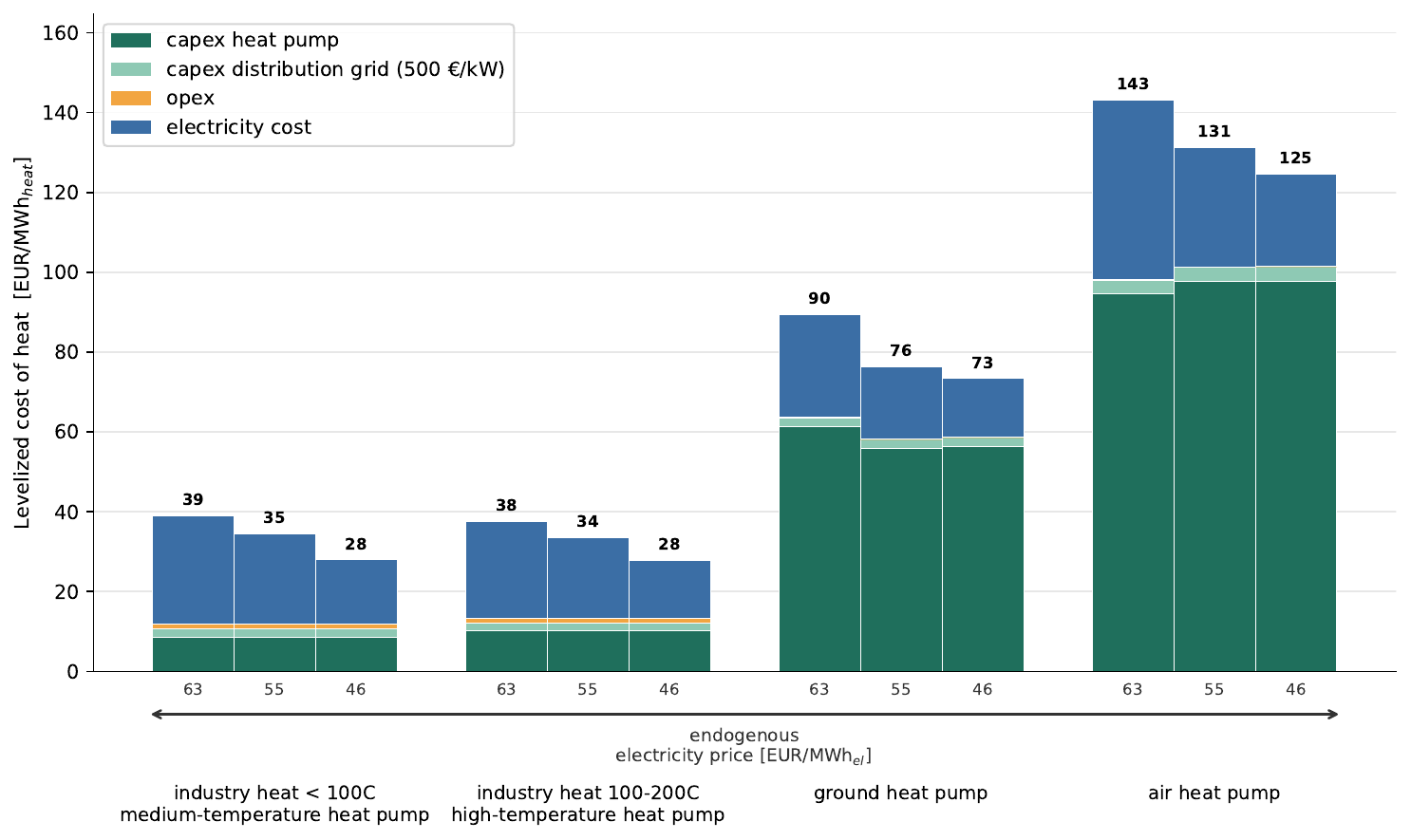}
    \caption{
      Disaggregation of heat pump LCOHs across supply technologies and model runs.
    }
    \label{fig:hp_lcoh_disaggregation}
\end{figure}

\begin{figure*}[t]
    \centering
    \includegraphics[width=0.8\textwidth]{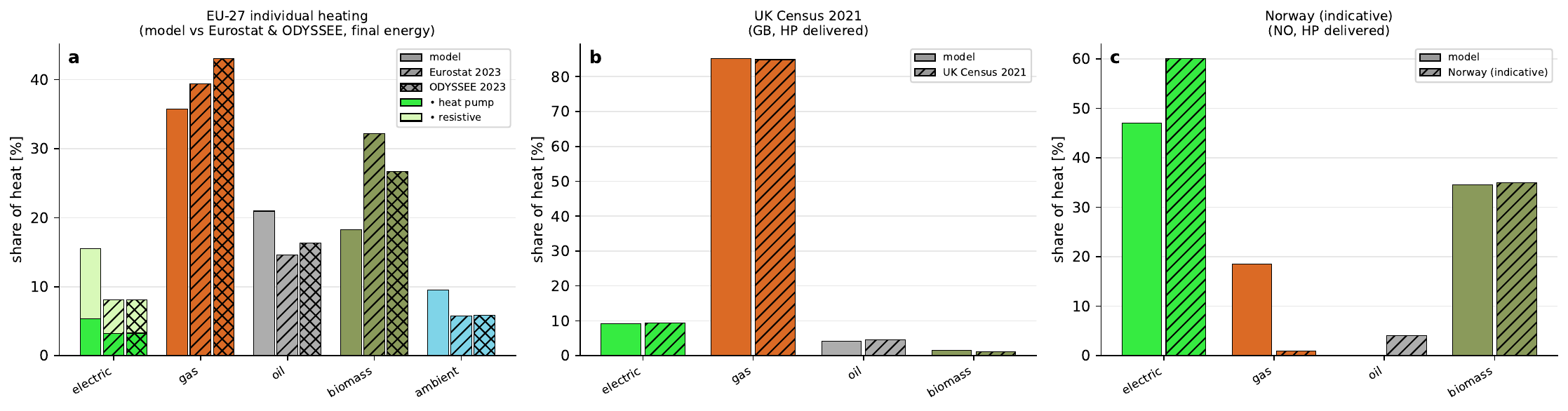}
    \caption{
      Comparison between system model (400$\,$bcm/a gas consumption) with brownfield existing heat capacities against 2023 heat supply from various sources. 
      The geographic scope varies between panels: we cover \textbf{a} EU27, \textbf{b} the United Kingdom and \textbf{c} Norway.
      Data sources: Eurostat \cite{eurostat2023households} and ODYSSEE-MURE \cite{odysseemure2023} for EU27 individual heating, EU district heating \cite{ecdh2017}, the UK Census 2021 \cite{ukcensus2021heating} and Energifakta Norge \cite{energifaktanorge}.
    }
    \label{fig:heat_validation}
\end{figure*}

\begin{figure}[t]
    \centering
    \includegraphics[width=.45\textwidth]{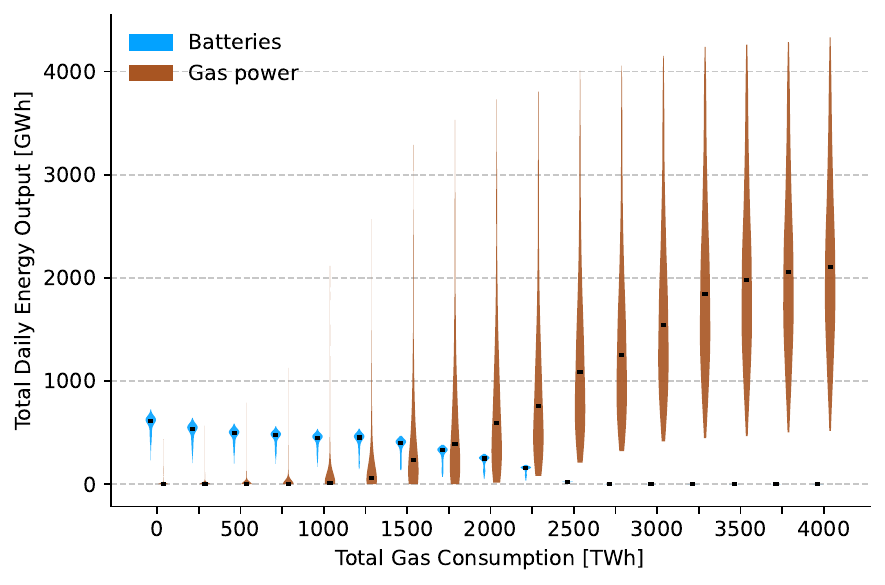}
    \caption{
      Distributions of daily power contributions (generation for gas turbines, discharging for batteries).
      The plot includes both utility-level and home batteries; gas turbines cover both CCGT and OCGT.
    }
    \label{fig:battery_gas_tradeoff}
\end{figure}

\begin{figure*}[t]
    \centering
    \includegraphics[width=\textwidth]{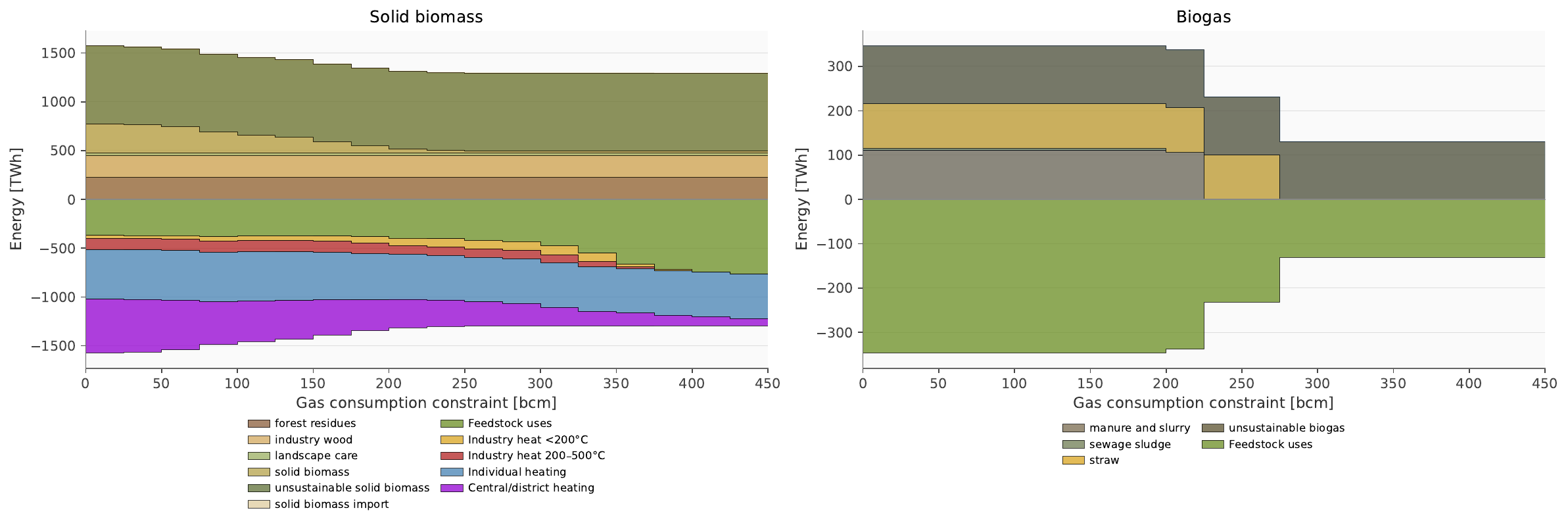}
    \caption{Solid biomass and biogas balances for different levels of gas consumptions.}
    \label{fig:total_biomass}
\end{figure*}

\begin{figure}[t]
    \centering
    \includegraphics[width=.45\textwidth]{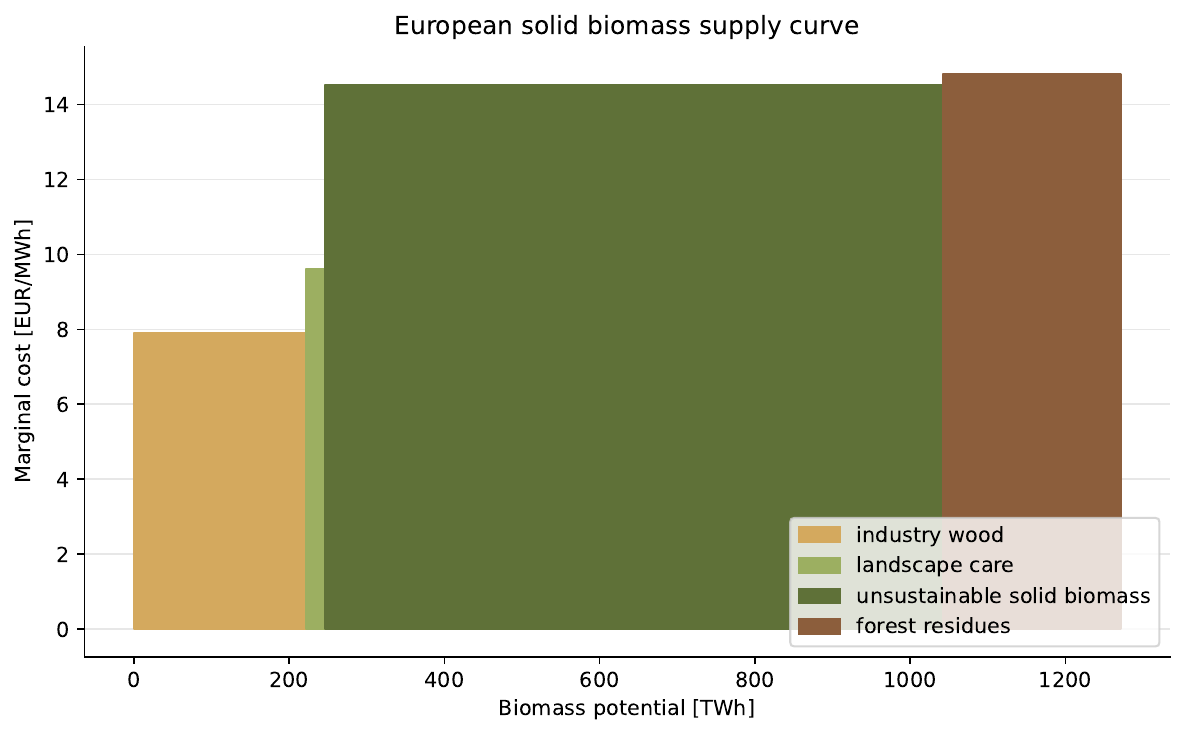}
    \caption{Supply curve in the model for solid biomass, taken from the Medium Potential of the ENSPRESO dataset \cite{ruiz_enspreso_2019}.}
    \label{fig:biomass_supply_curve}
\end{figure}

\begin{figure*}[t]
    \centering
    \includegraphics[width=.95\textwidth]{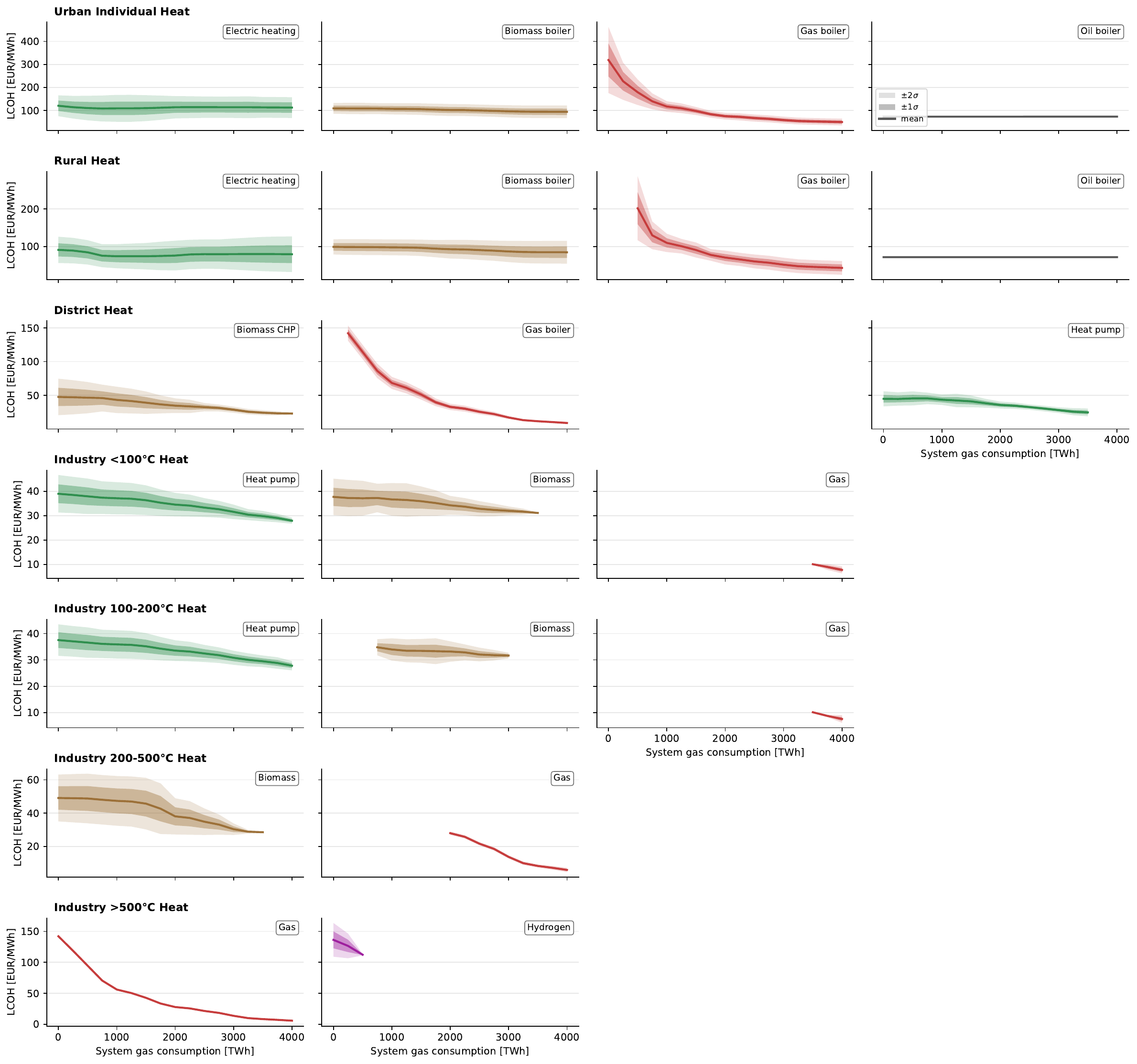}
    \caption{Levelised costs of heat (LCOH) in different heat demands in the model. The uncertainty bands refer to the spread across the 50 regions in the network. Data is omitted where a technology supplies less than 10 TWh of heat across the network.
    }
    \label{fig:lcohs}
\end{figure*}

\begin{figure*}[t]
    \centering
    \includegraphics[width=\textwidth]{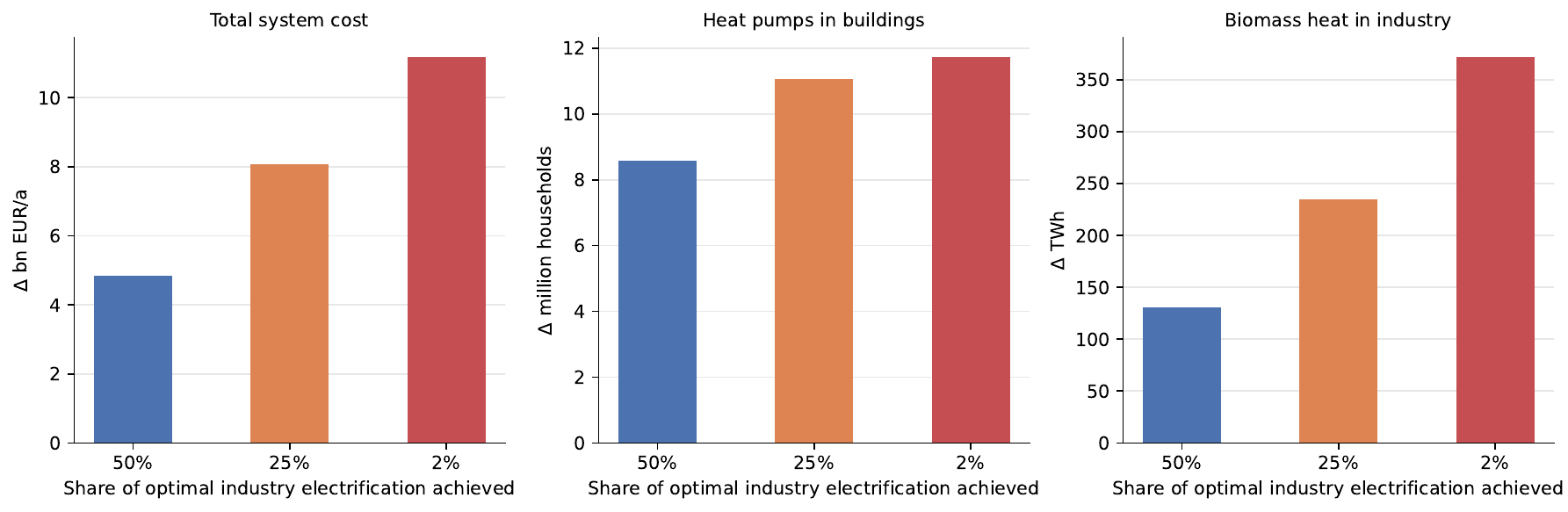}
    \caption{
      Impact on system cost, building heating and solid biomass use under 200 bcm gas use while constraining the share of industry heat that can be electrified.
      The share on the $x$-axis refers to the share of currently gas-met industry heat demand that can, in the respective model, be met by other means.
    }
    \label{fig:slow_industry_electrification_progress}
\end{figure*}

\begin{figure*}[t]
    \centering
    \includegraphics[width=\textwidth]{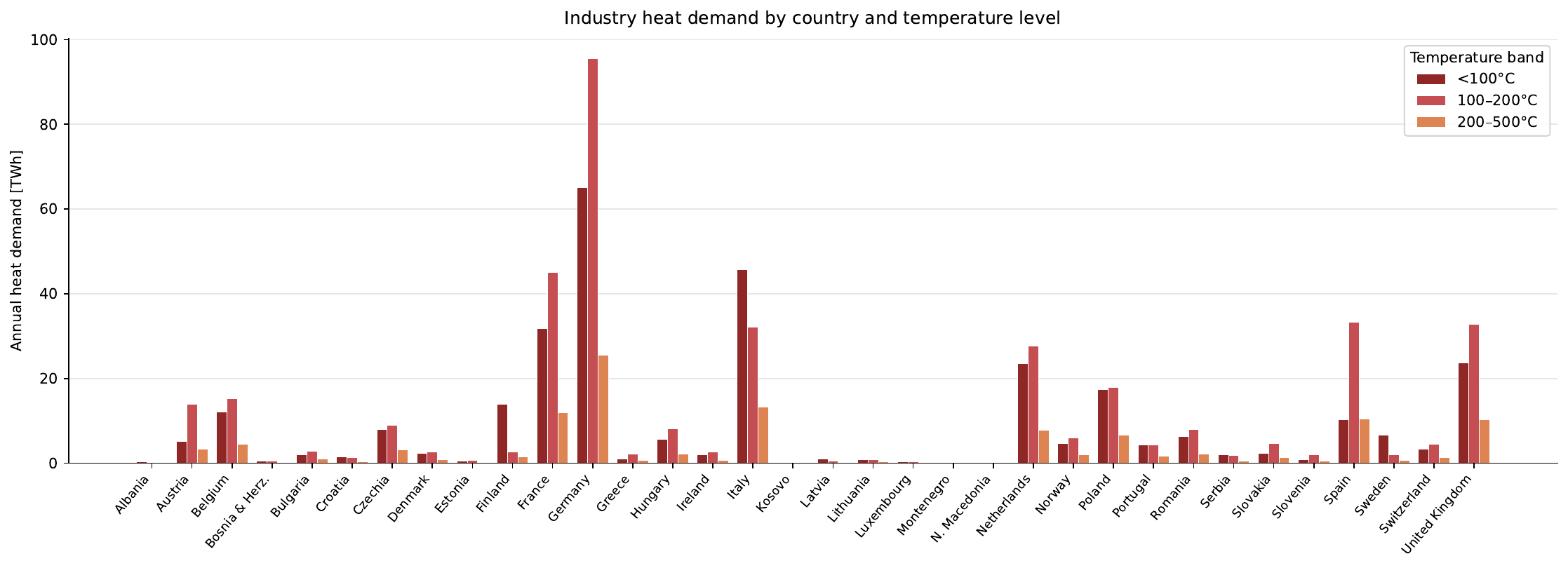}
    \caption{
      Industry heat demand below 500\deg C currently met by gas, by European country.
    }
    \label{fig:country_industry_heat_demand}
\end{figure*}

\begin{figure}[t]
    \centering
    \includegraphics[width=\linewidth]{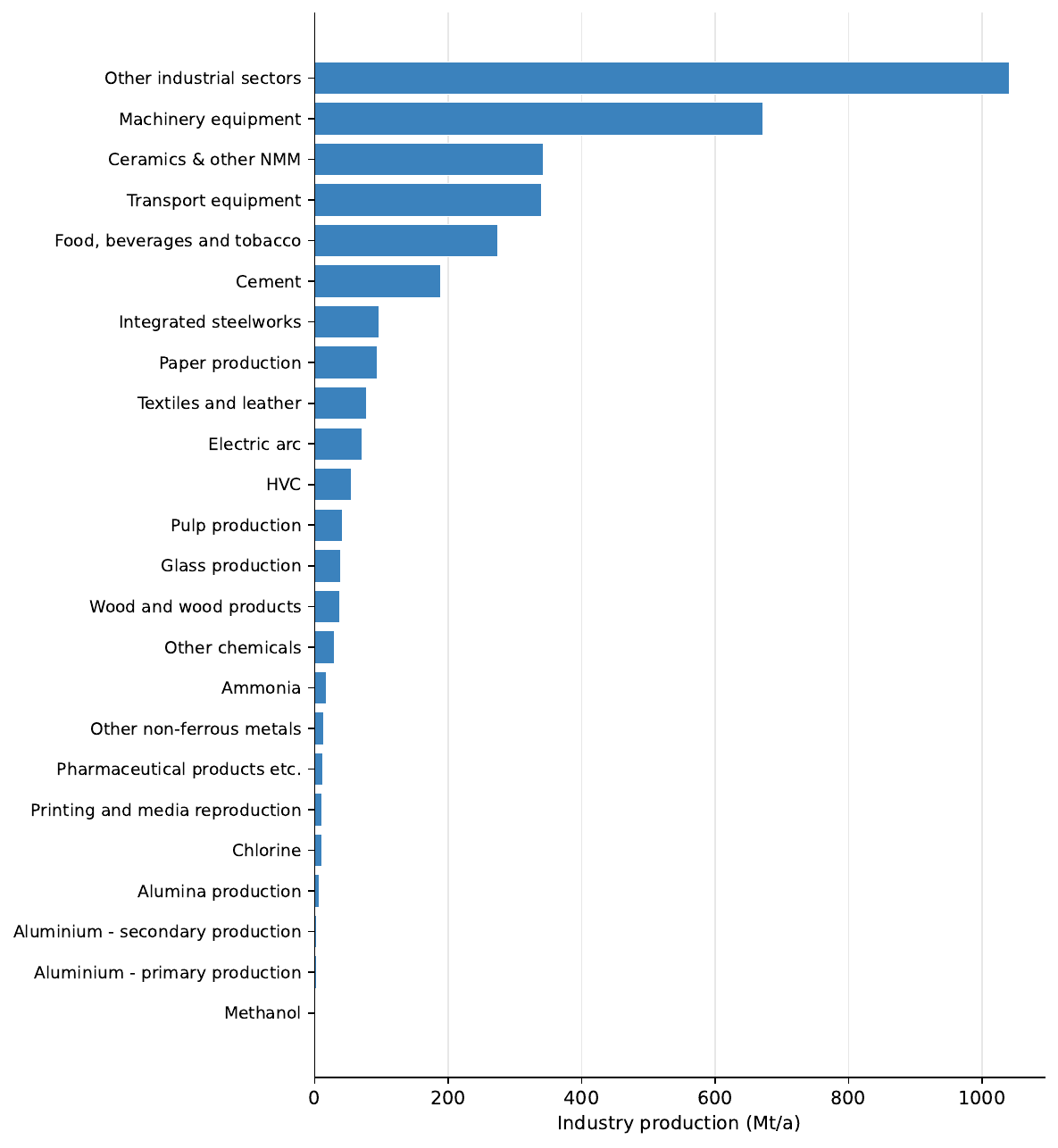}
    \caption{
      Annual production volumes for different industry sectors based on JRC IDEES dataset \cite{jrc_idees_2021}.
    }
    \label{fig:industry_sectors_production}
\end{figure}

\begin{figure}[t]
    \centering
    \includegraphics[width=\linewidth]{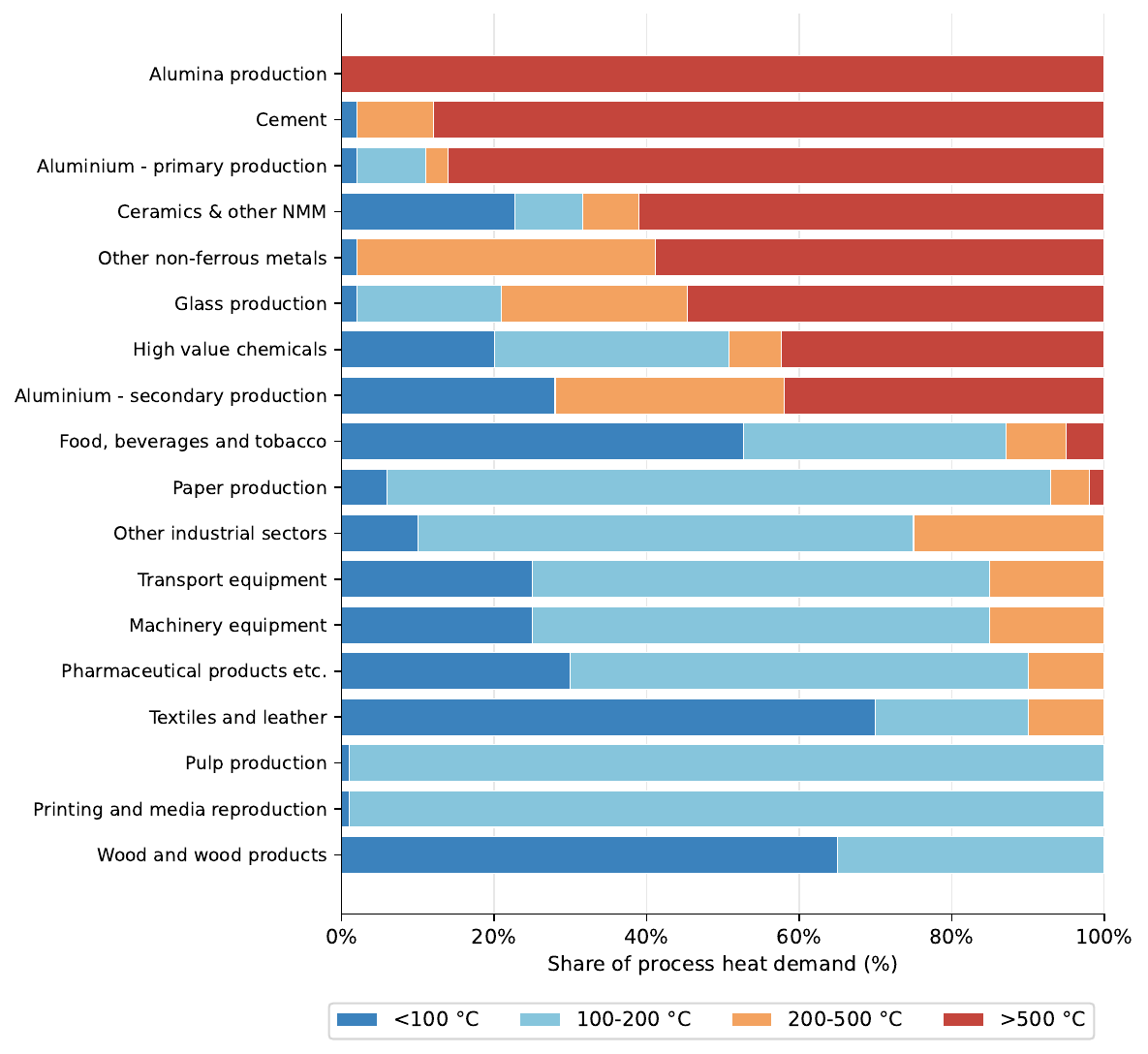}
    \caption{
      Share of process heat demand across the four temperature bands for each
      industrial sector, derived mostly from Fleiter et al.\ \cite{fleiter2025hydrogen}
      and supplemented for uncovered sectors as described in the main text.
      \textit{High value chemicals} refers to the basic petrochemicals,
      i.e.\ olefins (ethylene, propylene) and aromatics (benzene, toluene, xylene).
    }
    \label{fig:industry_temp_band_shares}
\end{figure}

\begin{figure}[t]
    \centering
    \includegraphics[width=\linewidth]{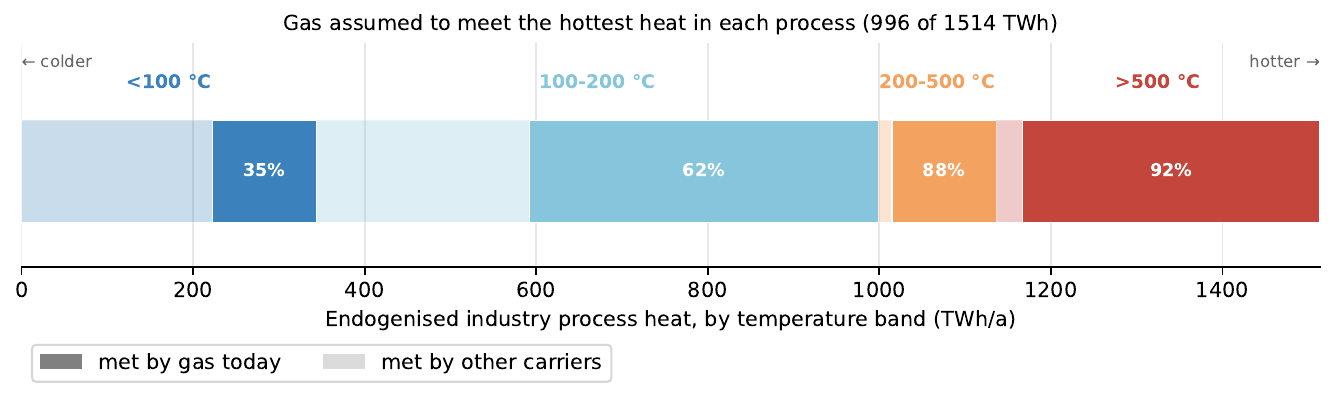}
    \caption{
      Visualisation of mapping from industry heat demands to gas demand.
      The JRC industry dataset shows that waste-heat, biomass and gas are used to supply heat.
      When an industry process utilises multiple of these, and also has heat demands in different temperature bands, then we assume it is the highest temperature-end of that heat demand that is met by gas since it is the only carrier that can provide higher temperature heat generically.
    }
    \label{fig:industry_hot_end}
\end{figure}

\begin{figure}[t]
    \centering
    \includegraphics[width=\linewidth]{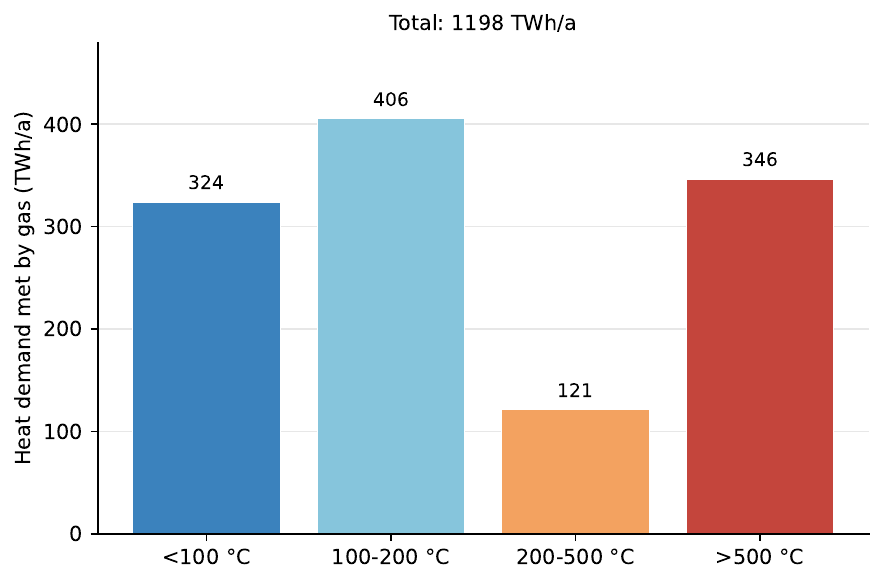}
    \caption{
      Gas-attributed industry heat demand per temperature band.
    }
    \label{fig:industry_heat_demand_per_band}
\end{figure}

\begin{figure*}[t]
    \centering
    \includegraphics[width=.95\textwidth]{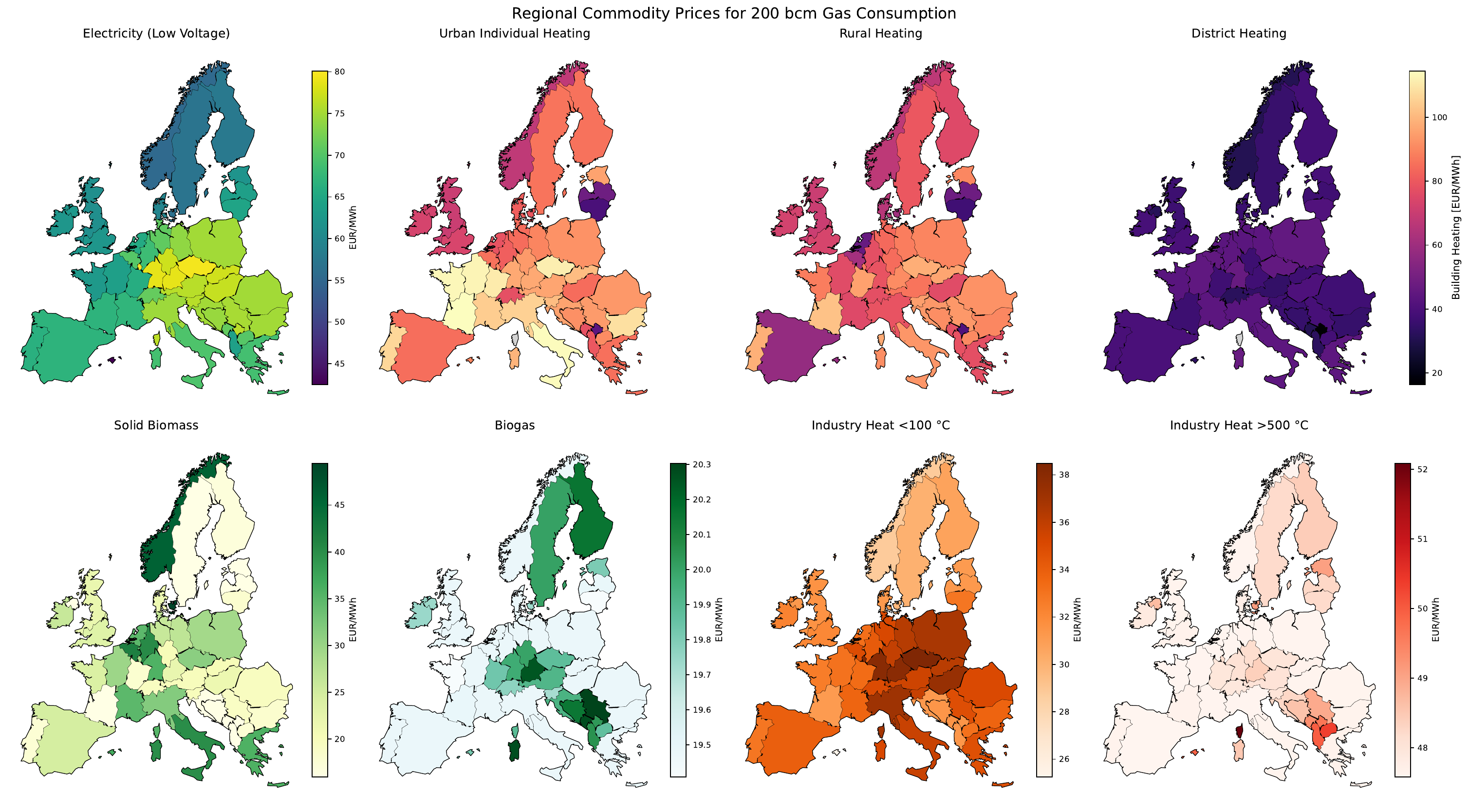}
    \caption{
      Time-weighted average marginal prices for different commodities under a 200$\,$bcm fossil gas supply constraint.
      [Higher prices in France in urban individual heating are somewhat misleading. It appears that price is largely driven by a few days of cold snaps with electric heating creating grid bottlenecks.]
  }
    \label{fig:regional_marginal_prices}
\end{figure*}

\begin{figure*}[t]
    \centering
    \includegraphics[width=.95\textwidth]{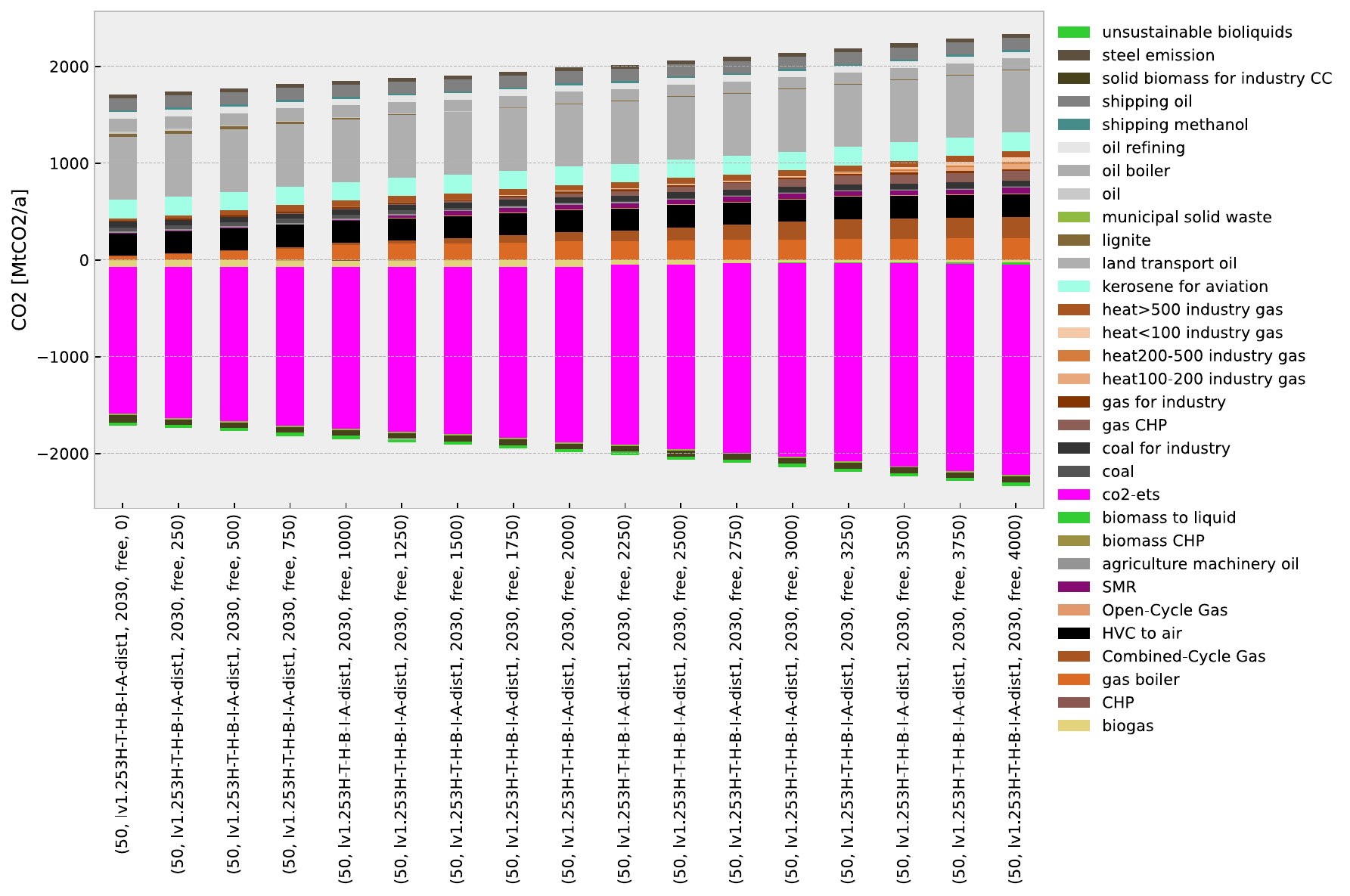}
    \caption{
      Composition of emissions under varying fossil gas supply constraints.
  }
    \label{fig:co2-balances}
\end{figure*}

\begin{figure*}[t]
    \centering
    \includegraphics[width=0.8\textwidth]{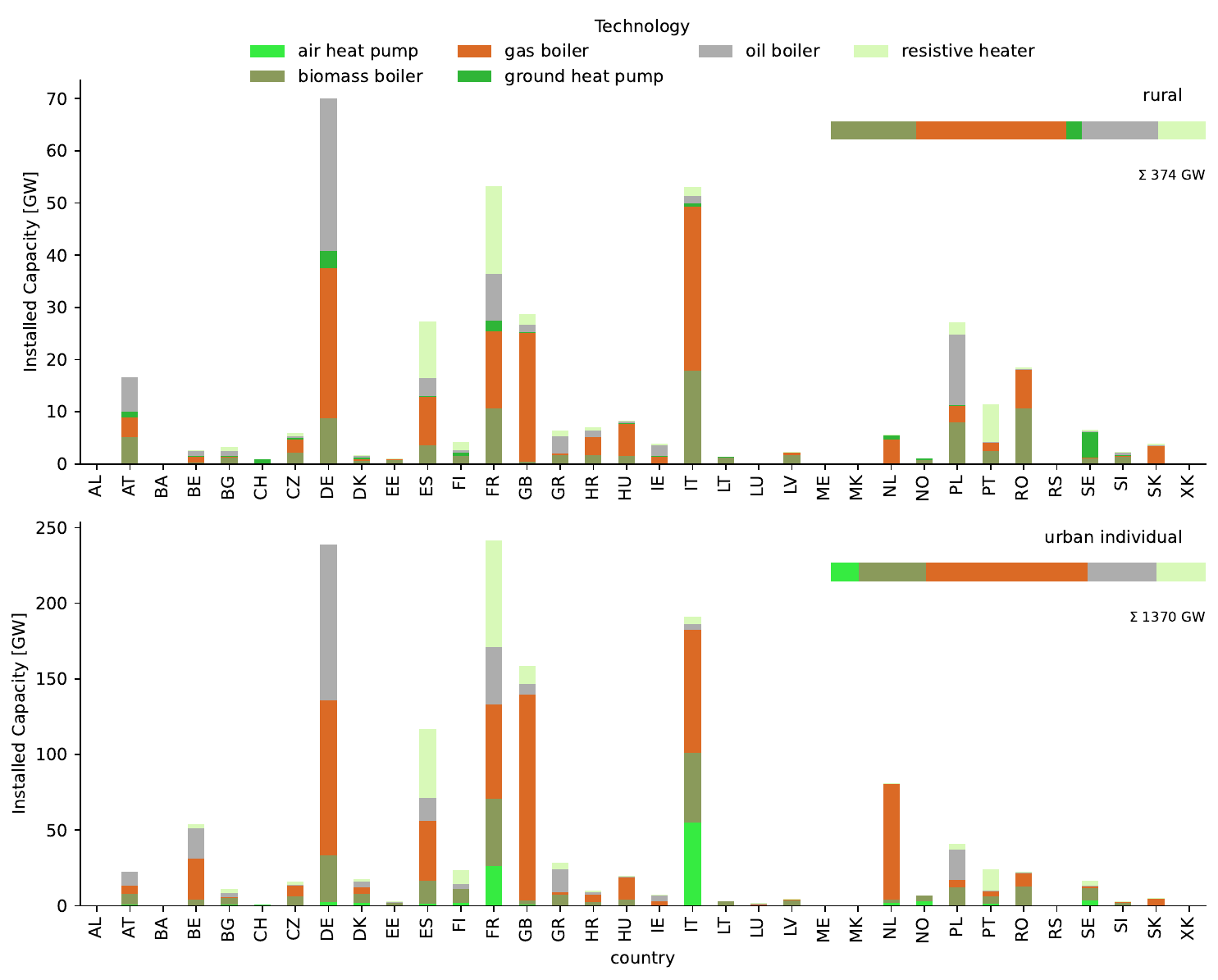}
    \caption{
      Installed heating capacities for all residential and service-sector heating, except district heating. The dataset used is from 2012 \cite{neumann2023potential}, suggesting that electric options are somewhat more widespread than represented here.
    }
    \label{fig:existing_heating_capacities}
\end{figure*}

\begin{figure*}
    \centering
    \includegraphics[width=1\linewidth]{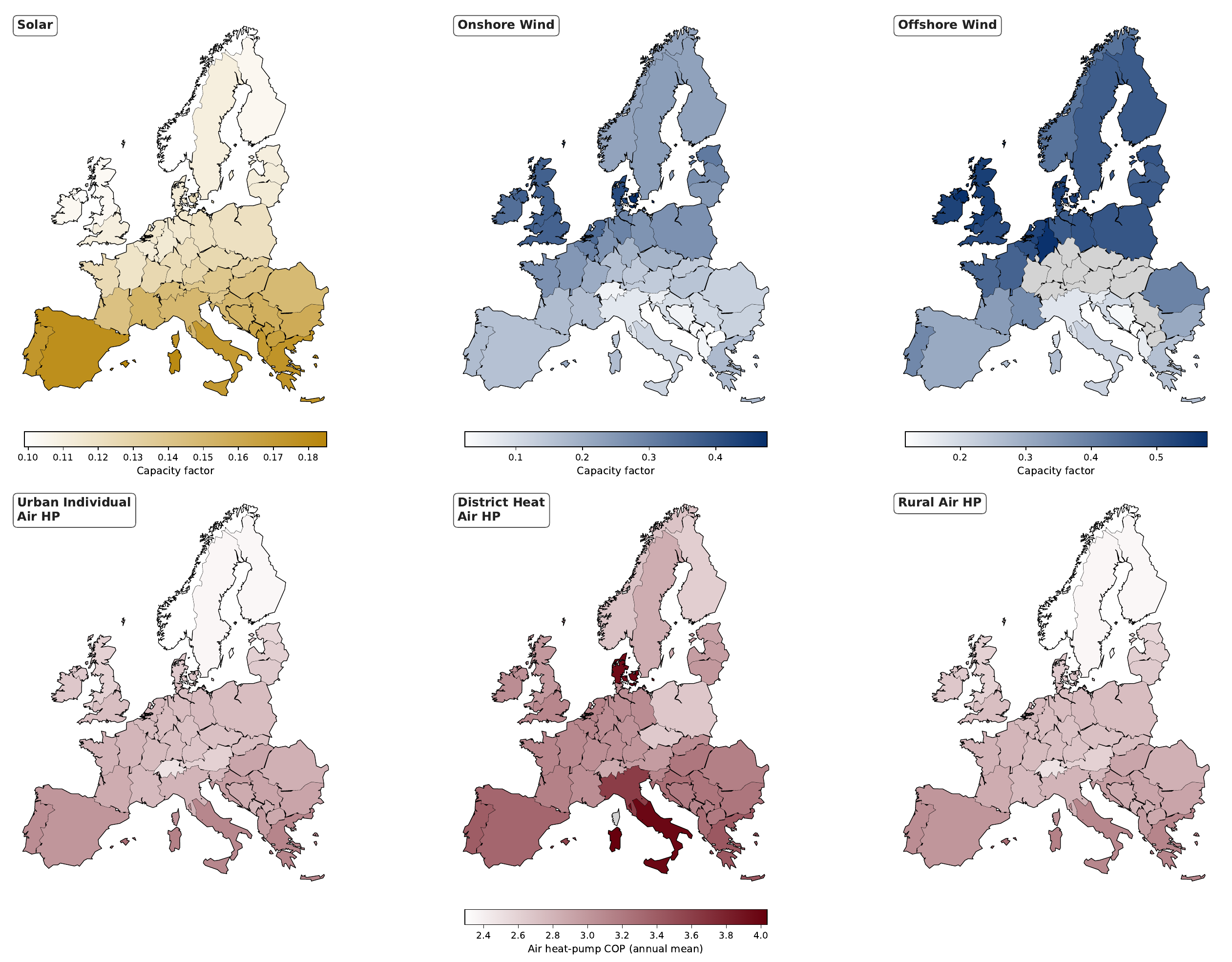}
    \caption{Capacity factors of renewables and, in the lower row, coefficients of performance for different kinds of heat pumps.}
    \label{fig:renewable_cf_maps}
\end{figure*}